\documentclass{article}

\usepackage{arxiv}
\usepackage[utf8x]{inputenc} 
\usepackage[T1]{fontenc}    
\usepackage{hyperref}       
\usepackage{url}            
\usepackage{booktabs}       
\usepackage{amsfonts}       
\usepackage{nicefrac}       
\usepackage{microtype}      
\usepackage{lipsum}
\usepackage{graphicx}
\usepackage[dvipsnames]{xcolor}
\usepackage{subcaption}
\usepackage{mwe}
\usepackage{amsmath}
\usepackage{multirow}
\usepackage{float}
\usepackage{bm}				
\usepackage{mathrsfs}
\usepackage{verbatim}

\title{Networked partisanship and framing: a socio-semantic network analysis of \\ the Italian debate on migration}

\author{
Tommaso Radicioni\thanks{Corresponding author: tommaso.radicioni@sns.it}\\
Scuola Normale Superiore\\
P.zza dei Cavalieri 7, 56126, Pisa (Italy)
\And
Tiziano Squartini
\\
IMT School for Advanced Studies\\
P.zza S. Ponziano 6, 55100, Lucca (Italy)
\And
Elena Pavan
\\
University of Trento\\
Via Verdi 26, 38122, Trento (Italy)
\And
Fabio Saracco
\\
IMT School for Advanced Studies\\
P.zza S. Ponziano 6, 55100, Lucca (Italy)\\
}
\begin{document}

\maketitle

\section*{Abstract}
The huge amount of data made available by the massive usage of social media has opened up the unprecedented possibility to carry out a data-driven study of political processes. While particular attention has been paid to phenomena like elite and mass polarization during online debates and echo-chambers formation, the interplay between online partisanship and framing practices, jointly sustaining adversarial dynamics, still remains overlooked. With the present paper, we carry out a socio-semantic analysis of the debate about migration policies observed on the Italian Twittersphere, across the period May-November 2019. As regards the social analysis, our methodology allows us to extract relevant information about the political orientation of the communities of users - hereby called \emph{partisan communities} - without resorting upon any external information. Remarkably, our community detection technique is sensitive enough to clearly highlight the dynamics characterizing the relationship among different political forces.As regards the semantic analysis, our networks of hashtags display a mesoscale structure organized in a core-periphery fashion, across the entire observation period. Taken altogether, our results point at different, yet overlapping, trajectories of conflict played out using migration issues as a backdrop. A first line opposes communities discussing substantively of migration to communities approaching this issue just to fuel hostility against political opponents; within the second line, a mechanism of distancing between partisan communities reflects shifting political alliances within the governmental coalition. Ultimately, our results contribute to shed light on the complexity of the Italian political context characterized by multiple poles of partisan alignment.

\section{Introduction}

Over the last two decades, an increasing number of studies have addressed the forms of digital public discourse and its implications for political processes. Particular emphasis has been put on \textit{adversarial dynamics} taking place online and echoing long-standing concerns on public discourse as a terrain of conflict and not only as the means through which it is expressed \cite{steinberg1998tilting}. In this respect, the contentious nature of digital discourse has been often connected to homophilic communicative interactions, which, in turn, have been seen as serving two opposed aims. On the one hand, the construction of collective counter-publics coalescing around alternative political narratives that challenge the stereotyping and the under-representation of women and people of color \cite{jackson2020hashtagactivism}; on the other hand, the fragmentation and tribalization of public debate, most notably within echo-chambers \cite{sunstein2001echo,10.1371/journal.pone.0159641} nurture the diffusion of disinformation \cite{tucker_guess_barbera_vaccari_siegel_sanovich_stukal_nyhan_2018,10.1371/journal.pone.0181821} and facilitate the polarization of political systems \cite{Conover2011}.

With specific reference to this latter strand of analysis, researchers pivoting around the study of political polarization have contributed to uncover the multidimensional, multilayered and variable nature of adversarial dynamics that take place in and through public discourse. More specifically, \cite{doi:https://doi.org/10.1002/9781118541555.wbiepc168} argues that progressive political polarization results from entwined processes of \emph{elite polarization} (i.e. patterns of progressive spacing between parties and party representatives compensated by increasing internal proximity) and \emph{mass polarization}, affecting citizens' decisions. However, a fundamental role is also played by partisan media, i.e. media outlets that aim at advancing peculiar political agendas \cite{10.2307/23496642}, and mainstream outlets \cite{arceneaux_johnson_2015,DIMAGGIO2013570,doi:10.1080/1354571X.2013.780348} whose contents tend to parallel those of political parties and leaders, thus fostering political fragmentation \cite{hallin_mancini_2004}.

Looking more closely at the digital context, authors in \cite{doi:10.1080/10584609.2020.1785067} remark that online polarized conversations entail interactional, ideological and affective elements. Altogether, these elements underpin the construction of antagonistic collective identities which are internally cohesive but increasingly far from those of political ``enemies''. Similarly, the exploration of online debates, with specific reference to the US context, shows that concentration of homophilic relations within segregated communities is driven by users' ideological orientation but also depends on the discussed topics. In fact, political conversations about electoral competition and societal policies have been shown to be remarkably more polarized than those about more ``ordinary'' topics \cite{BarberaEchoChamber}. Systematic investigations of online discussions about highly divisive topics, such as climate change \cite{WILLIAMS2015126} and migration issues \cite{fi12100173} further confirm the effects of segmented community structures, in terms of content circulation, already pointed out in \cite{10.1145/1134271.1134277, Conover2011}. As users tend to interact with other actors perceived as like-minded, different and divergent (when not directly polarized) interpretations of the same issue tend to emerge; as a consequence, the exposure to news and information becomes increasingly selective \cite{Schmidt3035}, thus reinforcing pre-existing ideological positions while, at the same time, marking deep fractures with political adversaries.

Situated understandings of reality that underpin and, at the same time, are nurtured by highly clustered networks of relations are found to relevantly affect voting and participation behaviors as well as opinions \cite{druckman_peterson_slothuus_2013}. Thus, as pointed out in \cite{doi:10.1177/1940161212474472}, contentious online dynamics often takes the form of a multidimensional competition that impacts how people will understand, remember and act upon an issue. Interestingly, as shown in \cite{doi:10.1177/0163443719876541}, competition between main political parties on Twitter occurs in digital contexts displaying variable levels of polarization, depending on countries' party system and electoral law arrangement (i.e., proportional or majoritarian). In this sense, besides ``perfectly polarized'' contexts like the US political system, a variety of conflicting political contexts can be found online that mix a two-pole antagonism with more articulated processes of community formation and content circulation \cite{fi12100173}.

In this paper, we aim at proposing an analytic framework to advance explorations of adversarial dynamics that take place in the digital space created by social media. In continuity with extant studies, we acknowledge the \emph{collective} and \emph{multiplex} nature of these dynamics as we conceive them as the result of entwined social and content-based networked processes animated by a variety of actors, such as citizens, journalists, political leaders, activists and civil society organizations. At the same time, we aim at emphasizing their \emph{multidimensionality} as homophilic patterns of relations that generate segregated and fragmented community. These online structures do, in fact, evolve over time in connection with major events occurring offline, both in the political arena and in the domain under examination. Investigations of online contentious dynamics pivoting around offline protest events highlight how conversations on social media exert a pre-mediation effect on on-site events. More specifically, several studies provided evidence of the relevance of digital communications for the spurring of offline protests \cite{doi:10.1080/1369118X.2015.1109697, Bastos_Recuero_Zago_2014, Jungherr2014} but also of the effects that events on the ground have for online discursive dynamics \cite{Pavan_Forthcoming}.

However, this multidimensional relationship is not always linear. For example, in \cite{https://doi.org/10.1111/jcom.12145} a certain variability is underlined in the causality between social media streams and offline protests while authors in \cite{fi12100173} observe that the levels of engagement in online discussions about migration issues in Italy and the actual distribution of migrants across the national territory are two independent variables.

With a particular view to understand more in detail the mechanisms that drive the fluid evolution of online adversarial dynamics, we propose to conceive online partisanship and framing as networked discursive processes through which attention is directed towards specific actors and particular frames are brought to prominence. Empirically, we translate our proposed approach into the two-fold exploration of social and semantic structures that are formed online. Focusing on the Twitter discussion about migration that emerged in Italy between May and November 2019, we explore networked partisanship dynamics by implementing a community-detection procedure that follows the approach outlined in \cite{Saracco_2019,Radicioni_2020}. Along these lines, we consider Twitter verified users as proxies of enlarged digital elites \cite{8750923,doi:10.1080/15405702.2016.1269909} and proceed to identify broader partisan communities that shape around them, following attention flows that pass through the retweeting activity of non-verified users. 

Additionally, we examine networked framing practices occurring within each partisan community by studying the semantic networks they induce via hashtagging practices, which we acknowledge to be powerful framing procedures \cite{doi:10.1177/1940161212474472,doi:10.1080/1369118X.2015.1109697,doi:10.1177/2158244015586000}. Furthermore, in order to account for the ever-evolving nature of 
online adversarial dynamics \cite{doi:10.1080/10584609.2020.1785067} while deepening their multidimensional nature, we perform our exploration according to a longitudinal perspective, by linking the topological study of our communities to main events occurring ``on the ground''. Our results further confirm the fluid nature of networked partisanship and framing while, at the same time, urging to consider a different rhythm along which their evolution is traced. Indeed, partisan and framing networked practices appear to co-evolve with the rapid sequence of both domain-specific events (particularly, controversial cases of search-and-rescue mission) and political events (particularly, changing political alliances cristallyzed, in our case study, by the governmental crisis in the summer of 2019). Moreover, our results point at different, yet overlapping, trajectories of conflict played out using migration issues as a backdrop. A first line opposes communities that discuss substantively of migration to those that approach this issue to fuel their hostility against political adversaries. Within this last type of communities, a second mechanism of distancing between partisan communities reflects shifting political alliances within governmental coalitions. Ultimately, our results contribute to shed light on the nature of the Italian political context which is a polarized environment characterized by a multipolar ideological system \cite{fi12100173, doi:10.1177/0163443719876541}.

\section{Materials and methods}

\subsection{Tweets collection}
\label{sec:data}

The starting point of our analysis is represented by tweets that have been publicly posted from 24 April 2019 to 24 November 2019. This period was heated by a set of relevant political events, as the European political elections (26 May 2019) and the Italian governmental crisis (end of August 2019) that ended up breaking the government coalition gathering the Five Stars Movement (\textit{Movimento 5 Stelle}) with the League party (\textit{Lega}) but also by a set of contested search-and-rescue operations performed by NGOs such as non-authorized entrance of the Sea Watch-3 boat in the Lampedusa port. A detaleid overview of of the most relevant events characterizing the period under analysis is present in \nameref{app:events} while in \cite{fi12100173} a more detailed description of the Italian debate about migration across the period August 2018-July 2019.

Tweets have been retrieved through the Twitter Streaming API and selected if containing at least one of the following hashtags: \emph{\#accoglienza} (`hospitality'), \emph{\#apriteiporti} (`open the ports'), \emph{\#chiudiamoiporti} (`close the ports'), \emph{\#immigrazione} (`immigration'), \emph{\#integrazione} (`integration'), \emph{\#migranti} (`migrants'), \emph{\#restiamoumani} (`let's stay human'), \emph{\#rifugiati} (`refugees'), \emph{\#sbarchi} (`disembarkation'), \emph{\#stopinvasione} (`stop the invasion').

The hashtags above have been chosen through a daily monitoring of Twitter trending topics having care to include a spectrum of positions in the controversy over migration. A summary of the most relevant hashtags used as anchors for our analysis is present in \nameref{app:hashtag_descriptions}. The data acquisition procedure led to a data set of approximately 5 million of tweets, posted by 306.894 users.

Hashtags contained in the tweets have then subjected to a pre-processing procedure where any two hashtags have been merged if found to be "similar" according to the Levenshtein (or edit) distance \cite{Survey_TextSimilarity}. Finally, for each couple of similar hashtags, only the most frequent has been considered in the final list. To merge only strings that are either typing errors or different conjugations of verbs/substantives (i.e., singular in place of plural and vice versa) we set a threshold to the maximum number of allowed differences between any two strings equal to 2.

Albeit our analysis seeks to cover the flow of contents and the interactions developed amongst users during the observation period, our work is not meant to provide an exhaustive portrayal of the entire Italian context: ad-hoc publics that assemble around topics, in fact, are not exhausted by communities that form on particular social media platforms - let alone around specific hashtags \cite{quteprints91812,hanna2013computer}; moreover, it is widely acknowledged that Twitter data systematically under-represent the real-world population \cite{HUBERTY2015992}. Nonetheless, our mapping of the Twitter discussion around migration provides a useful entry point to reason around processes of networked partisanship and framing. Indeed, the public assembled by the different anchor hashtags did engage in a `outright and deliberately public communication' \cite{quteprints91812} about migration issues, upon a platform that was not only widely diffused in Italy at that specific moment \cite{WeAreSocial} - according to Audiweb \cite{Vincos}, $\simeq 8$ millions of Italian users were active on Twitter in 2018 - but that also plays a pivotal political communication role \cite{Jungherr2016}, exerting a regular effect of agenda-setting on the country mainstream media \cite{doi:10.1080/17512786.2015.1040051}.

\subsection{Methods}
\label{sec:methods}

A bipartite network is an extremely versatile representation of the relationships between two disjoint set of nodes, also called \emph{layers}, where edges connect only nodes belonging to different sets. In mathematical terms, a bipartite network can be represented by a $N_\top\times N_\bot$ matrix $\mathbf{M}$, with $N_\top$ being the total number of nodes belonging to the first layer and $N_\bot$ being the total number of nodes belonging to the second layer.

The detection of the partisan communities and of the corresponding semantic networks require two distinct bipartite networks. First, we root the analysis of partisanship on a bipartite network containing two distinct set of Twitter accounts: the former is composed by users who are \emph{verified} by the platform (to guarantee that these accounts are `authentic, notable, and active' \cite{TwitterPolicies}) while the latter includes \emph{non-verified} accounts. In our mathematical representation, $m_{i\alpha}=1$ if user $i$ has retweeted user $\alpha$, or viceversa, at least once during the observation period - it is worth noticing that, in the selected Twitter data, a retweet is mainly performed by a non-verified account who share a tweet posted by a verified user. Second, we explore the framing processes starting from a bipartite network defined by the list of unique user IDs tweeting, or retweeting, the list of merged hashtags at least once. Hence, $m_{i\alpha}=1$ if the user $i$ has tweeted, or retweeted, at least once the hashtag $\alpha$ - and 0 otherwise. Beside computing an aggregate bipartite network that covers the whole observation period, we have also constructed monthly bipartite user-by-user and user-by-hashtag networks, to cope with the longitudinal unfolding of the conversations we monitored. It is worth noticing that the unweighted nature of our bipartite networks is motivated by the fact that the number of times an hashtag (or a verified user) is retweeted is not as relevant as the co-occurrence of that specific hashtag (or that verified user) with others, across the observation period.

Each bipartite network belonging to one of the two classes is, then, projected onto the layer of interest to obtain the corresponding monopartite network. Typically, monopartite networks are obtained in a na\"ive fashion, i.e. by linking any two nodes if the number of their common neighbors is found to be positive: ore quantitatively, given any two nodes $\alpha$ and $\beta$, such an algorithm prescribes to quantify the number of common neighbors as

\begin{eqnarray}
V_{\alpha\beta}^*=\sum_{j=1}^{N_\top}m_{\alpha j}m_{\beta j}
\end{eqnarray}
and connect them, in the corresponding monopartite projection, if $\Theta[V_{\alpha\beta}^*]=1$, i.e. if $V_{\alpha\beta}^*$ is strictly positive. Conversely, our analysis follows the approach outlined in \cite{Saracco_2017,Radicioni_2020}, according to which any two nodes are linked if $V_{\alpha\beta}^*$ is found to be statistically significant when compared against a properly-defined benchmark \cite{Saracco_2015}. Hereby, we employ a model belonging to the class of the Exponential Random Graphs (hereby, ERGs) and named Bipartite Configuration Model (BiCM) \cite{Saracco_2015,SquartiniCimini2019}. Thus, after computing the observed value $V_{\alpha\beta}^*$ for each couple of nodes $\alpha$ and $\beta$, the statistical significance of $V_{\alpha\beta}^*$ is quantified through the computation of a p-value, i.e.

\begin{eqnarray}
\text{p-value}(V_{\alpha\beta}^{*})=\sum_{V_{\alpha\beta}\geq V_{\alpha\beta}^{*}} f(V_{\alpha\beta})
\end{eqnarray}
where $f$ indicates the probability distribution of the values $V_{\alpha\beta}$ under the chosen null model: pairs of nodes are linked if the corresponding p-values are rejected by a multiple hypothesis testing procedure (see \nameref{app:inferring} for more details on this). The output of this second procedure is a $N_\bot\times N_\bot$ adjacency matrix $\mathbf{A}$ whose generic entry reads $a_{\alpha\beta}=1$ if nodes $\alpha$ and $\beta$ are part of a statistically significant number of V-motifs - and 0 otherwise. The procedure sketched above has been adopted for projecting both onto the layer of verified users and onto that of hashtags.

In the scientific literature, other validation procedures have been proposed. Examples of these techniques are provided by the method presented in \cite{10.1371/journal.pone.0017994}, where the authors combine a number of comparisons based upon the hypergeometric distribution with the False Discovery Rate (i.e., our validation procedure) for testing multiple hypotheses at a time, and \cite{10.1007/978-3-030-65347-7_55}, where this approach was implemented to study the 2019 Indonesian elections. As already observed in \cite{Saracco_2017}, where a comparison of different projection algorithms is carried out, the validation procedure employed in the present manuscript has been observed to be more effective than others in filtering a networked system.

\section{Results}
\label{sec:results}

\subsection{The social side: networked partisanship}

We start looking at networked partisanship by identifying \emph{discursive communities} of users that assemble around the use of migration-related hashtags \cite{Radicioni_2020}. In order to map these communities, we focus on a specific type of interaction amongst all the ones enabled by Twitter - i.e. retweets. While mentions and replies point to direct interactions with other users, retweets signal an explicit recognition (for better or for worse) of the contents published by a specific user. As such, retweets can be conceived of as a powerful `mode of repetition' \cite{doi:10.1177/1940161212474472} able to reinforce collective political identities \cite{Conover2011, doi:10.1080/14742837.2016.1268956}.

\paragraph{Identifying partisan communities.} Following the approach of \cite{Caldarelli2020,Becatti2019}, we built a bipartite network of 1.144 `verified' users and 115.885 `non-verified' users. The information about the identity of verified users, mainly figures of public interest (as politicians, newspapers and TV accounts) who have requested to be authenticated by Twitter, is automatically retrieved via Twitter APIs.

Networked partisanship dynamics are investigated by looking at the monopartite projection of the aforementioned bipartite network on the layer of verified users. Communities are, then, identified through a three-step procedure: 1) communities of verified users are isolated through the Louvain algorithm (see \nameref{Appendix_C} for more details on this); 2) several non-verified users are assigned to (one of) these communities, according to the highest scoring on a \emph{polarization index}, quantifying the level of `embeddedness' within each subgroup (see \nameref{Appendix_D} and \cite{Becatti2019} for further details); 3) the affiliation of the remaining non-verified users is inferred by `propagating' the initial community labels of both verified and polarized users in the retweeting network (see \nameref{Appendix_D} for further details).

\begin{figure}[!ht]
\centering
\includegraphics[width=0.8\textwidth]{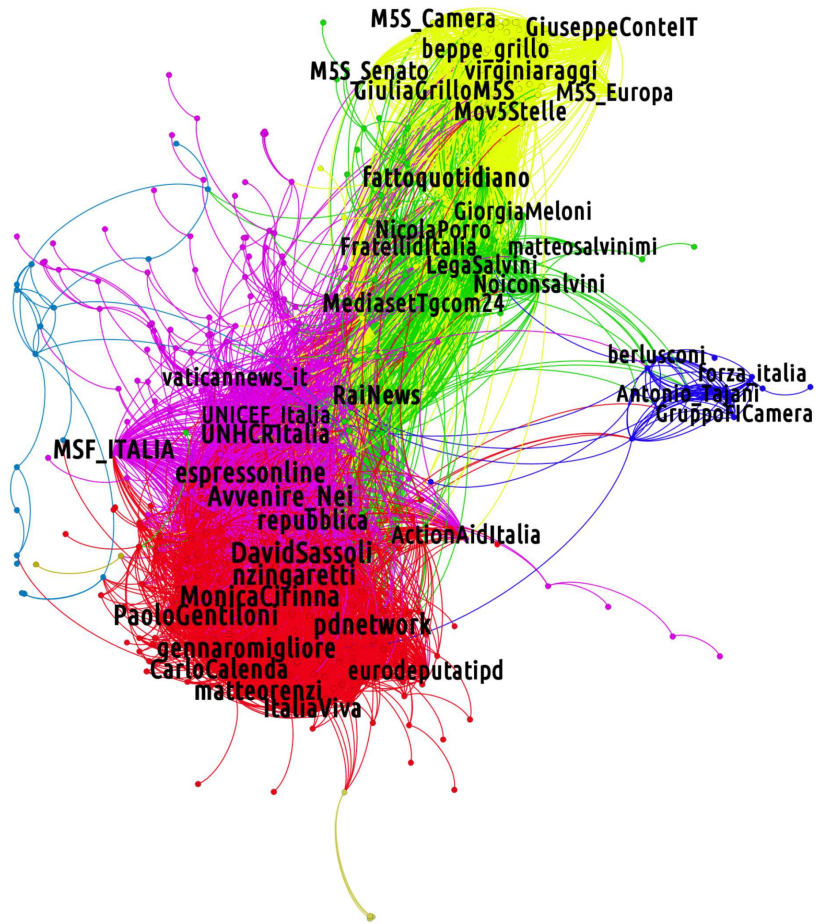}
\caption{{\bf Monopartite projection on the layer of verified users.} The network is obtained from the tweeting and retweeting activity of verified and non-verified users across the entire observation period (May-November 2019). The five final communities pivot around verified accounts of the main Italian political parties/coalitions and politicians (i.e. far-right parties as Brothers of Italy and the League, in green; center-left parties as the Democratic Party and Italy Alive, in red; the Five Stars Movement, in yellow; Go Italy in blue) as well as around accounts of media, intergovernmental and non-governmental organizations.}
\label{fig1}
\end{figure}

The monopartite projection on the layer of verified users is shown in Fig. \ref{fig1} and is further characterized by the communities revealed by the Louvain algorithm. The composition of each discursive community is, then, assessed by examining the verified users assigned to it. Remarkably, the detected clusters of users overlap with the main political parties/coalitions that were present in Italy after the 2018 Italian elections \cite{Radicioni_2020}. Through a manual check performed \textit{a posteriori}, the political affiliation of (approximately) all verified users, who are also members of Italian political parties, is correctly identified by the adopted method. In this sense, these discursive communities, induced by a common retweeting behavior, also configure themselves as \emph{partisan communities} where retweeting behaviors sustain identification mechanisms with one specific political array. These results are consistent with those of the analysis carried out independently in \cite{fi12100173} and further confirm the segregated, and partisan, nature of retweet networks already highlighted in \cite{Conover2011}.

More in detail, our procedure leads to identify five different partisan communities, engaged in discussions about migration issues:

\begin{itemize}
\item \textbf{Right-wing community (DX)} gathering official accounts of right-wing political parties such as Brothers of Italy (\textit{Fratelli d'Italia}) and the League (i.e., \textit{@FratellidItalia}, \textit{@LegaSalvini}), their leaders (i.e., \textit{@GiorgiaMeloni}, \textit{@matteosalvinimi}), politicians and journalists working for right-wing national newspapers (e.g. \textit{@NicolaPorro});

\item \textbf{Center-left wing community (CSX)}, gathering official accounts of the political parties composing the center-left alliance such as the Democratic Party and Italy Alive (i.e., \textit{Partito Democratico} and \textit{Italia Viva} with their accounts \textit{@pdnetwork} and \textit{@ItaliaViva}), their leaders (i.e. \textit{@nzingaretti}, \textit{@matteorenzi}), writers and journalists working for left-wing national magazines and newspapers (e.g. \textit{@eziomauro}, \textit{@robertosaviano});

\item \textbf{Five Stars Movement community (M5S)}, gathering accounts related to the Italian populist party named Five Stars Movement and including its leaders (e.g. \textit{@luigidimaio}, \textit{@Roberto\_Fico}) and institutional accounts (e.g. \textit{@M5S\_Camera}, \textit{@M5S\_Senato}) but also the national newspaper \textit{Il Fatto Quotidiano} and the journalists working for it (e.g. \textit{@fattoquotidiano}, \textit{@petergomezblog}). Notably, the Twitter account of the back-then Italian Prime Minister Giuseppe Conte (suggested by the Movement to overcome the impasse in forming a new government after the 2018 national elections) is included in this community;

\item\textbf{Go Italy community (FI)}, a smaller community gathering the official accounts of prominent members of the Go Italy party (\textit{Forza Italia}), initiated by Silvio Berlusconi (e.g. \textit{@berlusconi}, \textit{@gabrigiammarco}, \textit{@forzaitalia}); 

\item\textbf{Media, international governmental and non-governmental organizations community (MINGOs)}. While not strictly party-related, a fifth community emerges around a variety of verified accounts connected to three main sets of actors: 
\begin{itemize}
\item media as weekly national magazines such as \textit{L’Espresso} (\textit{@espressoline}), online newspapers like the Italian \textit{Huffington Post} or \textit{Il Post} (\textit{@huffpostitalia}, \textit{@ilpost}), television shows of investigative journalistic and documentary nature (\textit{@reportrai3}), journalists covering foreign affairs and migrations (e.g. \textit{@martaserafini}, \textit{@mannocchia});
\item accounts of non-governmental organizations aimed at defending human rights (e.g. \textit{@amnestyitalia}) or specialized on migration and international cooperation issues (e.g. \textit{@emergency\_ong}, \textit{@ActionAidItalia}), accounts of prominent activists in this domain (such as Regina Catrambon, the initiator of a search-and-rescue NGO called \textit{Migrant Offshore Aid Station}, and Carola Rackete, the captain of Sea-Watch 3, the ship entering the Italian port of Lampedusa regardless the opposition of the back-then Minister of the Internal Affairs Matteo Salvini);
\item accounts of international governmental organizations such as the Italian chapters of the UNICEF, the International Organization for Migrations and the United Nations Refugee Agency.
\end{itemize}
\end{itemize}

In Table \ref{tab:discursivecommunities} we display some properties of the partisan communities above: from the entire network of retweets, we consider the subgraph relative to each community and calculate few basic statistics. As it can be seen in Table \ref{tab:discursivecommunities}, the partisan community retweet subgraphs are characterized by structural differences, revealing different levels of activity. In terms of the number of users ($N_{u}$), the FI and the CSX communities are, respectively, the least and the most populated ones.

Notably, even if the number of users $N_u$ is pretty similar, the CSX and the DX communities display opposite behaviors: the average degree of the users within DX is nearly three times the one of the users within CSX - meaning that, on average, a DX user is three times more active than a CSX user; M5S accounts are also quite active, even if less than DX ones \cite{Caldarelli2020c}. For the sake of completeness, in Table \ref{tab:discursivecommunities} we also report the information regarding the normalized mean degree, i.e. $\langle k\rangle_N=\left\langle\frac{k-\min\{k\}}{\max\{k\}-\min\{k\}}\right\rangle=\frac{\langle k\rangle-\min\{k\}}{\max\{k\}-\min\{k\}}$. Since taking the average of the normalized degrees is equivalent at normalizing the mean degree itself, the behavior of $\langle k\rangle_N$ also provides information about the range of variation of the degrees: as Table \ref{tab:discursivecommunities} reveals, the degrees of DX users are, overall, more similar (i.e. their range of variation is smaller) than the degrees of the users of the other communities.

\begin{table}[!ht]
\centering
\caption{
{\bf Structural characteristics of the five partisan communities.} Partisan communities show distinct structural characteristics - e.g. the number of users $N_{u}$, number of edges $N_{e}$, mean degree $\langle k\rangle$ and normalized mean degree $\langle k\rangle_{N}$ - but exhibit a rather similar communicative behavior in terms of the polarization index $\rho_{\alpha}$ and the self-reference indexes for retweeting ($\mu_r$) and mentioning ($\mu_m$).}
\begin{tabular}{c|c|c|c|c|c|c|c} 
\hline
Partisan community & $N_{u}$ & $N_{e}$ & $\langle k\rangle$ & $\langle k\rangle_{N}$ & $\langle\rho_{\alpha}\rangle$ & $\mu_r$ & $\mu_m$ \\ [0.5ex] 
\hline
DX & 70061 & 2075013 & 59.2 & 4$\cdot$10$^{-4}$ & 0.95 & 0.93 & 0.52 \\ [0.5ex]
CSX & 112459 & 1080271 & 19.2 & 9$\cdot$10$^{-4}$ & 0.95 & 0.9 & 0.47 \\ [0.5ex]
M5S & 6313 & 107100 & 33.9 & 7$\cdot$10$^{-3}$ & 0.84 & 0.73 & 0.39 \\ [0.5ex]
FI & 598 & 3133 & 10.5 & 1$\cdot$10$^{-2}$ & 0.89 & 0.77 & 0.34 \\ [0.5ex]
MINGOs & 24994 & 65388 & 5.2 & 6$\cdot$10$^{-4}$ & 0.93 & 0.78 & 0.47 \\ [0.5ex]
\hline
\end{tabular}
\label{tab:discursivecommunities}
\end{table}

Looking more closely at the type of interactions sustained by networked partisanship, a rather regular pattern seems to emerge. The overall community structure appears to be highly segregated, as the parameter $\langle\rho_{\alpha}\rangle$ (see \nameref{Appendix_D} for more details on this) reveals: in fact, almost the totality of the neighbors of each community members tends to be part of that same community, as already observed in \cite{Becatti2019} and \cite{fi12100173}. Moreover, all communities endorse the same communicative behavior as their users tend to employ retweets to broadcast opinions and contents generated by their own members while mentions establish an indirect contact with users in other communities. This element emerges by looking at the self-reference indexes $\mu_r$ and $\mu_m$, which are calculated as the ratio of the number of retweets ($\mu_r$) and mentions ($\mu_m$) of the users belonging to a given community and the total number of retweets, or mentions, performed within that same community. Self-reference indexes proxy the degree of internal and external influence of users, within and across communities, via their retweeting and mentioning activities, as in \cite{Cha10measuringuser} where a slightly different definition of $\mu_r$ and $\mu_m$ is employed. While the self-reference index $\mu_r$ shows that retweets tend to be employed to re-broadcast contents produced `internally', thus underpinning the formation of partisan collective identities, the values of $\mu_m$ reveals that, albeit separated, these communities are characterized by some levels of inter-activity.

Against this common background, there is nonetheless space for topological variation. A deeper cleavage seems to separate the largest DX and CSX communities from the rest of the discussion, as these two communities show highest levels of segregation $\langle\rho_\alpha\rangle$ and highly self-referential communicative behaviors $\mu_r$. Conversely, the other three communities show a higher tendency to broadcast internally also contents produced elsewhere.

\begin{figure}[!ht]
\centering
\includegraphics[width=\textwidth]{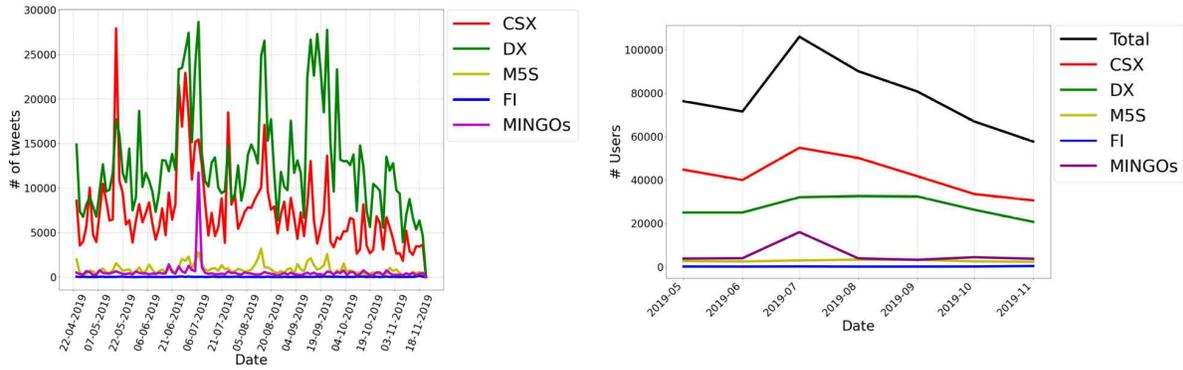}
\caption{{\bf Evolution of the number of tweets and users within partisan communities across the entire observation period.} In general, the trend of tweets (on the left) is characterized by weekly oscillations where peaks, coinciding with relevant political or issue-related events (e.g. the 2019 European elections, the ‘Sea-Watch 3’ crisis, the Italian government crisis) appear. The community producing the highest number of tweets is the DX one, followed by the CSX and the M5S ones. In correspondence of July 2019, a peak characterizing the trend of the number of users (on the right) within each partisan community is clearly visible due to the `Sea-Watch 3' controversy, which also induces a single community of `governmental supporters'.}
\label{fig2}
\end{figure}

\paragraph{Analyzing the rhythm of partisan communities.} As shown in Fig. \ref{fig2}, the level of activity within the five communities is characterized by weekly oscillations and is relevantly affected by events taking place at specific points in time: the 2019 European elections (end of May 2019), the Sea-Watch 3 episode (end of June 2019), the Italian government crisis (end of August - beginning of September 2019). Similarly to what has emerged from previous analyses focused on the Italian case \cite{Radicioni_2020,Caldarelli2020c,fi12100173} the right-wing community maintains higher (re)tweeting volumes than others. The CSX and M5S communities present a systematically lower number of tweets - except for few isolated peaks of activity characterizing the CSX community.

\paragraph{Attention flows within and between partisan communities.} In order to trace the attention flows within and between partisan communities, mentions and retweets are detected for all tweets posted by their members and the most mentioned and retweeted accounts are identified.

Overall, the list of retweeted users reflects the political affiliation of each partisan community, further confirming the value of this specific interaction feature for the construction of partisan collective identities. However, the results in Table \ref{tab:mention_retweet} reveal two different modes of constructing these identities. On the one side, the DX, M5S and FI communities display an institutional pattern as most retweeted accounts belong to either parties or political leaders. On the other side, the most retweeted accounts by the CSX and, even to a larger extent, the MINGOs communities belong either to media or non-governmental organizations. For example, the most retweeted account in the CSX community refers to the online newspaper \textit{Linkiesta}, the second one to the `Caritas Italiana', a catholic NGO based in Milan; other accounts belong to public figures very active on social media, as Roberto Saviano, a journalist well-known for his reports about mafia crimes in Southern Italy. As for the MINGOs community, the presence of catholic organizations reveals an attention towards the account of the Pope (whose pleas to solidarity resonated loud during the observation period), media outlets active in this area (like the catholic newspaper \textit{L'Avvenire}), etc.

\begin{table}[!ht]
\centering
\caption{
{\bf Five most retweeted and mentioned verified accounts for each partisan community.} While retweets are assumed to represent a broadcast action, mentions are indicative of an interactive relationship \cite{Conover2011}. Account names with an asterisk are intended as external to the partisan community.}
\resizebox{\textwidth}{!}{\begin{tabular}{cc|cc|cc|cc|cc}
\hline
\multicolumn{2}{c|}{\textbf{DX}} & \multicolumn{2}{c|}{\textbf{M5S}} & \multicolumn{2}{c|}{\textbf{CSX}} & \multicolumn{2}{|c|}{\textbf{FI}} & \multicolumn{2}{c}{\textbf{MINGOs}}\\
\hline
{\it Retweets} & {\it Mentions} & {\it Retweets} & {\it Mentions} & {\it Retweets} & {\it Mentions} & {\it Retweets} & {\it Mentions} & {\it Retweets} & {\it Mentions}\\
\hline
matteosalvinimi & matteosalvinimi & fattoquotidiano & matteosalvinimi$^{*}$ & caritas\_milano & matteosalvinimi$^{*}$ & renatobrunetta & forza\_italia & Pontifex & repubblica$^{*}$ \\
GiorgiaMeloni & GiuseppeConteIT$^{*}$ & virginiaraggi & GiuseppeConteIT & Linkiesta & repubblica & msgelmini & matteosalvinimi$^{*}$ & repubblica$^{*}$ & matteosalvinimi$^{*}$ \\ 
LegaSalvini & repubblica$^{*}$ & GiuseppeConteIT & luigidimaio & robertosaviano & GiorgiaMeloni$^{*}$ & berlusconi & berlusconi & UNHCRItalia & LaStampa$^{*}$ \\ 
Capezzone & GiorgiaMeloni & Mov5Stelle & virginiaraggi & repubblica & pdnetwork & GiorgioBergesio & simonebaldelli & Pontifex\_it & Avvenire\_Nei \\ 
NicolaPorro & luigidimaio$^{*}$ & carlosibilia & Mov5Stelle & CarloCalenda & CarloCalenda & mara\_carfagna$^{*}$ & renatobrunetta & Avvenire\_Nei & RaiNews \\ 
\hline
\end{tabular}}
\label{tab:mention_retweet}
\end{table}

Conversely, mentions are transversely used to interact (also) with members of other communities. Nonetheless, results in Table \ref{tab:mention_retweet} better specify the substance of cross-community interactions and confirm that discursive dynamics accompany and, to some extent, overlap with offline political alliances. While the DX and the M5S communities reciprocally open up to each other as they jointly sit in the government chaired by Giuseppe Conte, the other communities all find in Matteo Salvini a common target for their communications. This common trends toward addressing directly the account of the back-then Minister of the Internal Affairs, strongly positioned against migration and hostile to search-and-rescue missions as well as to sheltering operations, results in a strong personalization of the overall debate on migration, which ends up pivoting around the role and the responsibility of this specific individual.

These findings highlights how social media contribute to polarization processes: while discursive communities coalesce via retweets, they also interact, often in an adversarial way, via direct mentions. In order to complement these findings, in \nameref{Appendix_E} we quantify the social influence of users via the Hirsch index (or h-index, see \cite{Brexit2017}): the results further confirms the centrality of Matteo Salvini who is the verified users with the greatest value of h-index.



\begin{figure}[!ht]
\centering
\includegraphics[width=0.9\textwidth]{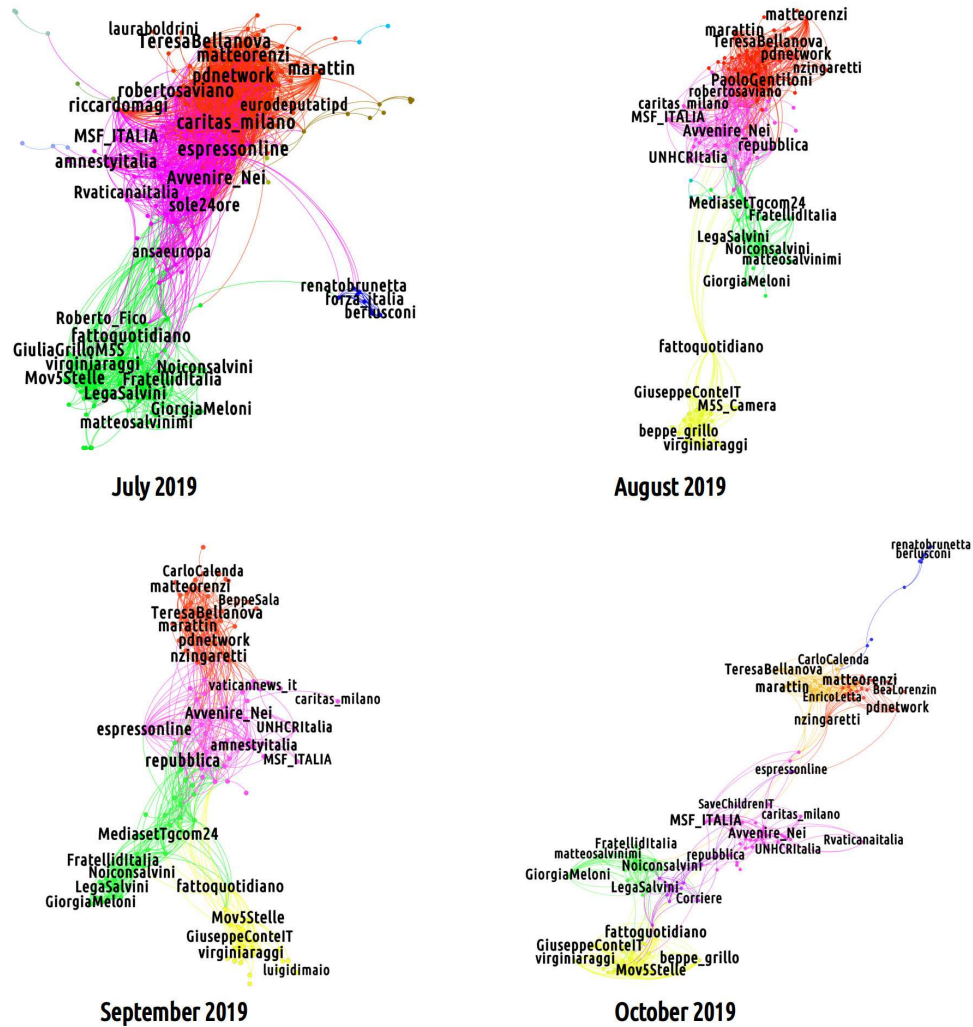}
\caption{{\bf Evolution of the partisan communities at a monthly time scale from July to October.} In October 2019, politicians of the main center-left party (united in the red cluster in September) split into two sub-communities, the orange one being induced by the Twitter activity of the members of a new political formation (i.e. the Italy Alive party).}
\label{fig3}
\end{figure}

\paragraph{Networked partisanship over time.} In order to examine the evolution of networked partisanship over time, we partitioned the data set on a monthly scale and defined a set of bipartite networks by considering the retweeting activity of non-verified users across these limited observation periods only. Close inspection of monthly attention flows allows us to grasp the evolution of discursive alliances mostly within (but also across) partisan communities in tight connection with the main events and political dynamics taking place "offline". Two examples well illustrate this point. As shown in Fig. \ref{fig3}, the 2019 Italian government crisis between August and September 2019 triggers a fracture in the CSX community which reflects the internal breaking of the Democratic Party: in particular, some users change their retweeting behavior, revealing the birth of a new partisan community (in orange in Fig. \ref{fig3}) pivoting around accounts connected with Italy Alive, the party founded by former Prime Minister Matteo Renzi in overt opposition to the Democratic Party.

A similar trend can be outlined for the right-wing sector. Albeit sharing seats in the Italian government, the Five Stars Movement and the League party are supported online by two systematically separated partisan communities. At the same time, while not tied to any common government responsibility, the League and the right-wing Brothers of Italy are merged within the same partisan community. While this misalignment speaks to the existence of a fracture within the governmental coalition \cite{oner2020growing}, networked partisanship dynamics evolve fluidly and, in some occasion, overcomes disagreements internal to the governmental coalition. On July 2019 the two M5S and DX communities merge after the entrance, without permission, of the rescue boat Sea-Watch 3 into the Italian territorial waters and the arrest of its captain, Carola Rackete. This event, occurred at end of June 2019, inflamed the political debate and forced governmental parties to overcome (at least formally) deeper disagreements and present itself as a united coalition. At this point, online partisan communities revolving around the two governmental parties also blended, signaling the progressive, albeit transient, formation of a discursive coalition supporting the government line.

\subsection{The semantic side: networked framing}

Let us now study the networked framing of the Italian debate about migration by considering the topological features of the semantic networks induced by each partisan community.

For each community, we constructed monthly user-hashtag bipartite networks and projected them on the layer of hashtags, via the procedure described in the \nameref{sec:methods} section and in \nameref{app:inferring}. Consistently with previous studies, we centered our attention on hashtags acknowledging that they "create visibility for a message [\dots] not only marking context but also changing and adding content to the tweet" \cite{doi:10.1177/2158244015586000}. More specifically, we consider hashtags as key devices to enact networked framing practices within networked publics \cite{doi:10.1177/1940161212474472,doi:10.1080/1369118X.2015.1109697,doi:10.1177/2056305118764437,jackson2020hashtagactivism} and, following Recuero et al. \cite{doi:10.1177/2158244015586000}, recognize their association within tweets as a strategy to convey specific narratives but also to mobilize specific audiences (see \nameref{app:hashtag_descriptions} for more details on this.)

\paragraph{Identifying conductive hashtags.} In order to identify the most relevant hashtags representing topics-, actors-, events-related slogans or references within the Twitter discussion, we have calculated their betweenness centrality which, in semantic terms, can be considered a proxy for the conductivity of a concept \cite{https://doi.org/10.1111/j.1468-2885.1993.tb00070.x}. More formally, betweenness quantifies the percentage of shortest paths passing through each hashtag and can be calculated as

\begin{equation}
b_\gamma=\sum_{\beta(\neq\alpha)}\sum_\alpha\frac{\sigma^{\alpha\beta}_\gamma}{\sigma^{\alpha\beta}}
\end{equation}

where $\sigma^{\alpha\beta}_\gamma$ is the number of shortest paths between hashtags $\alpha$ and $\beta$ passing through hashtag $\gamma$ and $\sigma^{\alpha\beta}$ is the total number of shortest paths between hashtags $\alpha$ and $\beta$. Then, the values of betweenness are normalized by the factor $(N-1)(N-2)/2$ (where $N$ is the number of hashtags within the specific semantic network) in order to compare the values of our centrality measure on networks of different size.

The comparison between most conductive hashtags for each community confirms the results on networked partisanship: as Table \ref{tab:centrality_1} shows, the hashtag \textit{\#salvini} is steadily among the first three hashtags in both the CSX and DX communities, across the entire observation period - hence, also after Salvini's exclusion from office In other words, the two communities engage in discussions that personalizes the debate on migration, making the controversial figure of Matteo Salvini central also from a semantic perspective. Interestingly enough, as shown also in \cite{fi12100173}, Salvini steadily appears among the top positions since August 2018, approximately one year before our observation period.

\begin{table}[!ht]
\centering
\caption{{\bf Ranking of the ten most central hashtags (according to their values of betweenness centrality) within the DX and the CSX partisan communities.} While some of the DX community hashtags as \textit{\#portichiusi} (`closed ports') denotes the clear anti-migration position of such a community, the CSX community is characterised by slogans as \textit{\#portiaperti} (`open ports') openly promoting pro-migration positions. The normalized betweenness of each hashtag, multiplied by a factor equal to $10^8$, is reported in parentheses.}
\resizebox{\textwidth}{!}{\begin{tabular}{c|c|c|c|c|c|c}
\hline
\multicolumn{7}{c}{\textbf{DX}} \\ 
\hline
2019-05      & 2019-06                      & 2019-07       & 2019-08             & 2019-09      & 2019-10        & 2019-11      \\
\hline
salvini (12)     & salvini (6.0)     & salvini (1.2)       & openarms (4.5)    & salvini (4.4)            & salvini (2.9)     & salvini (5.7)     \\
italia (4.4)      & fico (2.6)        & lampedusa (1.0)     & salvini (4.3)     & italia (3.1)             & malta (2.8)       & italia (4.5)      \\
governo (3.5)     & italia (2.3)      & seawatch3 (1.0)     & lampedusa (2.2)   & ong (3.1)                & italia (2.6)      & ong (3.2)         \\
libia (3.2)       & libia (2.3)       & italia (0.9)        & ong (2.1)         & portichiusi (2.6)        & lampedusa (2.0)   & lamorgese (3.1)   \\
portichiusi (2.9) & ong (2.3)         & ong (0.9)           & italia (2.1)      & dalleparoleaifatti (2.1) & oceanviking (1.8) & libia (2.9)       \\
m5s (2.6)         & lampedusa (2.1)   & francia (0.8)       & bibbiano (1.8)    & pd (2.0)                 & lamorgese (1.7)   & lega (2.2)        \\
pd (2.5)          & portichiusi (2.0) & bibbiano (0.7)      & portichiusi (1.6) & migrante (1.9)           & conte (1.1)       & governo (2.1)     \\
seawatch3 (2.3)   & pd (1.8)          & carolarackete (0.7) & pd (1.5)          & conte (1.9)              & giustizia (1.1)   & conte (1.8)       \\
marejonio (2.3)   & europa (1.7)      & seawatch (0.7)      & richardgere (1.3) & lampedusa (1.8)          & trieste (1.0)     & italiani (1.8)    \\
europa (2.0)      & giustizia (1.6)   & libia (0.7)         & gregoretti (1.0)  & macron (1.8)             & mattarella (0.9)  & oceanviking (1.7) \\
\hline
\multicolumn{7}{c}{\textbf{CSX}} \\ 
\hline
2019-05      & 2019-06                      & 2019-07       & 2019-08             & 2019-09      & 2019-10        & 2019-11      \\
\hline
salvini (2.6)      & salvini (2.4)                      & carolarackete (1.6) & openarms (3.3)            & libia (3.8)        & lampedusa (3.3)      & italia (2.9)       \\
facciamorete (1.8) & libia (2.3)                        & seawatch (1.5)      & salvini (2.8)             & marejonio (2.5)    & libia (1.5)          & libia (2.5)        \\
europa (1.1)       & giornatamondialedelrifugiato (1.6) & salvini (1.3)       & decretosicurezzabis (2.1) & salvini (2.4)      & erdogan (1.2)        & oceanviking (1.7)  \\
libia (1.0)        & lavoro (1.5)                       & seawatch3 (1.3)     & facciamorete (1.4)        & oceanviking (2.0)  & salvini (1.1)        & lavoro (1.7)       \\
milano (1.0)       & italia (1.3)                       & italia (1.2)        & lampedusa (1.3)           & lavoro (1.9)       & italia (1.1)         & facciamorete (1.0) \\
seawatch (1.0)     & seawatch (1.3)                     & libia (0.9)         & gregoretti (1.0)          & europa (1.7)       & europa (1.0)         & sicurezza (0.9)    \\
sicurezza (0.8)    & facciamorete (1.3)                 & lampedusa (0.9)     & libia (0.9)               & lampedusa (1.6)    & 3ottobre (1.0)       & salvini (0.7)      \\
marejonio (0.8)    & innovazione (0.9)                  & facciamorete (0.9)  & oceanviking (0.8)         & facciamorete (1.5) & malta (0.9)          & formazione (0.7)   \\
lampedusa (0.8)    & roma (0.9)                         & mediterranea (0.8)  & ong (0.7)                 & italia (1.4)       & sostenibilit√† (0.7) & portiaperti (0.6)  \\
lavoro (0.7)       & lampedusa (0.8)                    & ong (0.8)           & ai (0.6)                  & malta (1.3)        & cloud (0.6)          & europa (0.6) \\
\hline
\end{tabular}}
\label{tab:centrality_1}
\end{table}

While, at first sight, this result may suggest a semantic alignment between the two communities, a closer look at other conductive hashtags helps specifying that the common reference to Salvini provides the baseline for setting on divergent positions with respect to migration issues. This is well exemplified by hashtags recalling political slogans: while in the DX community the hashtag \textit{\#portichiusi} (`closed ports') remains present until the falling of the League-Five Stars Movement government, the CSX community repeatedly invites to collective actions via the hashtag \textit{\#facciamorete} (`let's act as a network') and adopts the counter-hashtag \textit{\#portiaperti} (`open ports') after the Democratic Party replaces the League in the second government led by Giuseppe Conte.

The hashtags displayed in Table \ref{tab:centrality_2} allow us to gain insight into the semantic positioning of the other two main partisan communities. Interestingly enough, consistently with its swinging governmental alliances, the M5S community appears to be "semantically torn". On the one hand, users in this community direct attention towards calls to collective action coming from the CSX community, as shown by the transversal adoption of the hashtag \textit{\#facciamorete}, and claim for the liberation of captain Rackete (\textit{\#freecarola}). On the other, they claim to "stop immigration" (\textit{\#stopimmigrazione}), supporting Salvini's positions on the topic and raising concern against the Democratic Party - particularly after its involvement into a (supposed) scandal about minors foster-care permits in the city of Bibbiano (\textit{\#bibbiano}). Taken altogether, these elements suggest that the M5S community does not hold a position as polarized as that of the DX and the CSX partisan groups on migration issues.

\begin{table}[!ht]
\centering
\caption{
{\bf Ranking of the ten most central hashtags (according to their values of betweenness centrality) within the M5S and the MINGOs partisan communities.} The M5S community presents a peculiar mix of hashtags characterizing both the CSX, as \textit{\#facciamorete} (`let's act as a network'), and the DX community, as \textit{\#bibbiano}, along with original hashtags referring to the governmental crisis in August 2019 and the formation of a new government in September 2019, as \textit{\#governoconte2} (`government Conte 2') and \textit{\#salvinitraditore} (`Salvini traitor'); on the other hand, the MINGOs community is characterized by an evident support towards specific pro-migration topics and slogans, as hashtags like \textit{\#ioaccolgo} (`I host') prove The normalized betweenness of each hashtag, multiplied by a factor equal to $10^8$, is reported in parentheses.}
\resizebox{\textwidth}{!}{\begin{tabular}{c|c|c|c|c|c|c}
\hline
\multicolumn{7}{c}{\textbf{M5S}} \\ 
\hline
2019-05      & 2019-06                      & 2019-07       & 2019-08             & 2019-09      & 2019-10        & 2019-11      \\
\hline
ricerca (159)                  & salvini (450)        & salvini (1010)      & facciamorete (73)    & governo (383)         & quota100 (515)       & pattoperlaricerca (868) \\
salvini (152)                  & disastrocalenda (446) & pd (913)          & m5s (44)             & governoconte2 (139)    & scuola (506)         & governo (330)          \\
m5s (83)                      & m5s (398)            & vonderleyen (858)  & italia (36)          & conte (109)           & turchia (405)        & libia (310)            \\
seawatch3 (83)                & piazzapulita (398)   & fico (850)        & salvini (32)         & lega (102)            & salvini (252)        & bellanova (211)        \\
bergamo (68)                  & seawatch3 (358)         & freecarola (780)   & boldrini (27)        & salvini (94)         & malta (233)          & zaiadimettiti (147)    \\
skytg24 (59)                  & governo (354)        & democrazia (772)  & lega (26.9)            & legammerda (75)      & disperati (178)      & leggedibilancio (147)   \\
governodelcambiamento (52)    & lampedusa (353)      & facciamorete  (582) & decretosicurezza (25) & m5s (75)             & giulemanidaroma (174) & lega (145)             \\
berlusconi (51)               & 21giugno (306)       & bibbiano (548)    & crisidigoverno (25)   & lamorgese (69)       & lega (172)           & salvinivergognati (143) \\
precari (50)                  & salviniusa (300)     & quota100 (449)    & salvinicazzaro (24.8)  & stopimmigrazione (65) & precari (142)        & salvini (136) \\
roma (49.4)                     & dimaio (281)         & ong (432)         & salvinitraditore (21.8) & oceanviking (63)     & 10ottobre (134)      & lamorgese (122) \\  
\hline
\multicolumn{7}{c}{\textbf{MINGOs}} \\ 
\hline
2019-05      & 2019-06                      & 2019-07       & 2019-08             & 2019-09      & 2019-10        & 2019-11      \\
\hline
libia (54)         & giornatamondialedelrifugiato (91) & seawatch3 (34)     & openarms (165)            & europa (181)        & lampedusa (73)     & libia (45)             \\
europa (49)        & papafrancesco (71)                & libia (28)         & libia (81)               & libia (181)         & 3ottobre (43)      & italia (31)            \\
lavoro (44)        & libia (67)                        & lampedusa (26)     & lampedusa (74)           & lampedusa (96)     & migrants (20)      & 31ottobre (30)         \\
papafrancesco (32) & inclusione (55)                   & carolarackete (20) & europa (61)              & rohingya (95)      & manovra (18)       & europa (28)            \\
1maggio (27)       & sostenibilità  (53)               & papafrancesco (18) & rohingya (55)            & mediterraneo (83)  & welfare (15)       & lavoro (19)            \\
salvini (26)       & lampedusa (52)                    & seawatch (17)      & gregoretti (45)          & oceanviking (75)   & europa (15)        & papafrancesco (16)     \\
lampedusa (19)     & italia (50)                       & italia (16)        & mediterraneo (44)        & italia (73)        & 15ottobre (14)     & caporalato (13)        \\
opportunità  (18)  & ioaccolgo (49)                    & rohingya (16)      & decretosicurezzabis (38) & marejonio (70)     & libia (13)         & 15novembre (13)        \\
rohingya (18)      & europa (42)                       & europa (16)        & salvini (37)             & conte (61)         & papafrancesco (12) & corridoiumanitari (12) \\
marejonio (18)     & rohingya (41)                     & 26giugno (15)      & tunisia (32)             & papafrancesco (61) & gruppohera (12)    & oceanviking (8.3) \\ 
\hline
\end{tabular}}
\label{tab:centrality_2}
\end{table}

However, the M5S community starts distancing itself from the DX community since the government crisis in August 2019: the relevance of this political event for the M5S is shown by the presence of hashtags that refer to the crisis, as \textit{\#crisidigoverno} (`government crisis') and \textit{\#governoconte2} (`government Conte 2'). Moreover, as the Five Starts Movements negotiates with the Democratic Party to set up a new governmental coalition, users of the M5S community put an increasing semantic distance between themselves, the League and Matteo Salvini, using hashtags calling him a `traitor' (as shown by the hashtag \textit{\#salvinitraditore}).

The MINGOs community, instead, appears to be semantically focused on the issue of migration and admittedly far from internal political matters. Due to the high presence of NGOs specialized on human rights and migration issues, this community takes a strong pro-migration stance, through hashtags such as \textit{\#ioaccolgo} (`I host') and \textit{\#inclusione} (`inclusiveness'). Similarly, the presence of catholic organizations orients the terms of the discussion around the figure of the Pope who is semantically recalled as a positive figure, opposed to that of Salvini.

\begin{figure}[!ht]
\centering
\includegraphics[width=\textwidth]{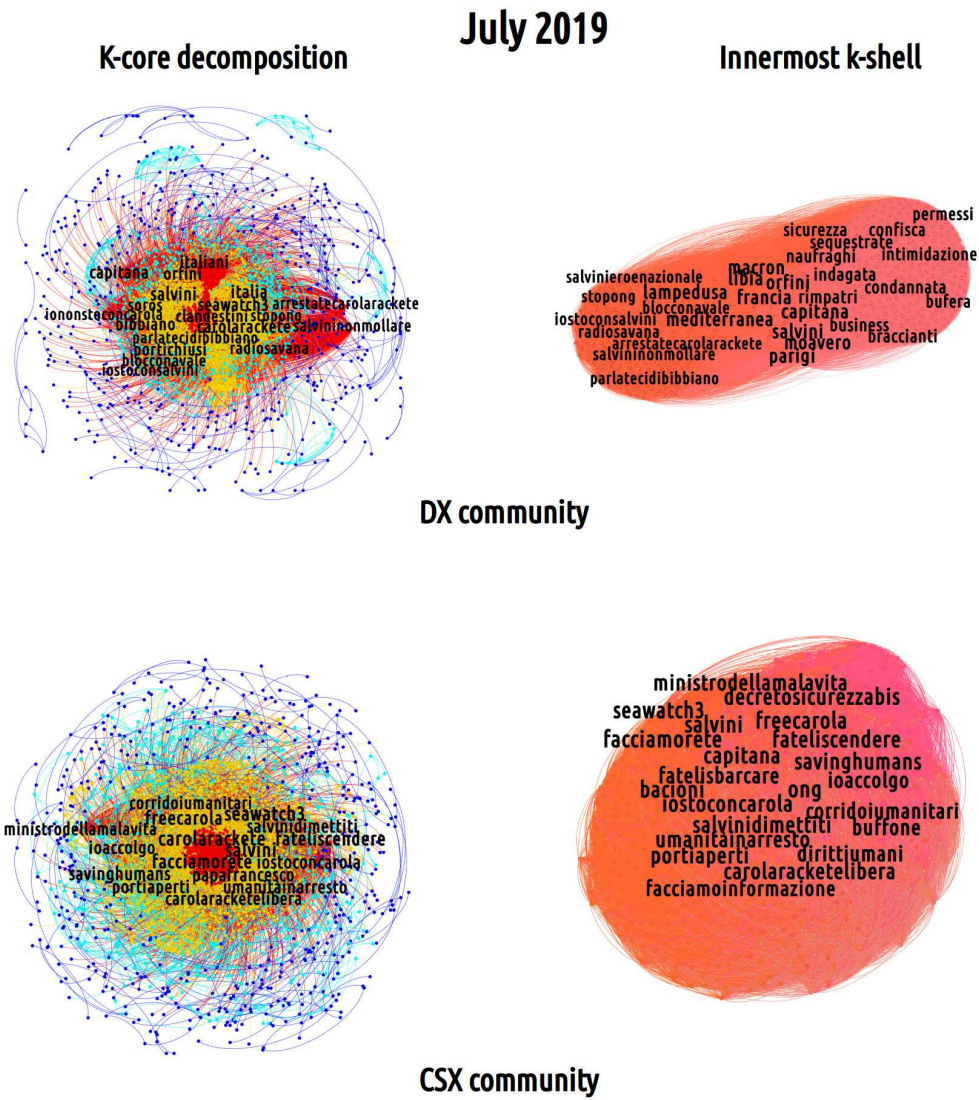}
\caption{{\bf K-core decomposition of the July 2019 semantic networks for the DX (top left) and the CSX (bottom left) partisan communities.} The k-core decomposition reveals the bulk of the discussion about immigration developed by these two partisan communities: while the innermost core of the DX-induced semantic network (top right figure) is composed by hashtags like \textit{\#salvininonmollare} (`Salvini don't give up'), \textit{\#arrestatecarolarackete} (`arrest Carola Rackete'), \textit{\#iostoconsalvini} (`I stand with Salvini'), that of the CSX-induced semantic network (bottom right figure) is composed by hashtags like \textit{\#salvinidimettiti} (`Salvini resign'), \textit{\#fateliscendere} (`let them get off'), \textit{\#carolaracketelibera} (`free Carola Rackete').}
\label{fig5}
\end{figure}

\begin{figure}[!ht]
\centering
\includegraphics[width=\textwidth]{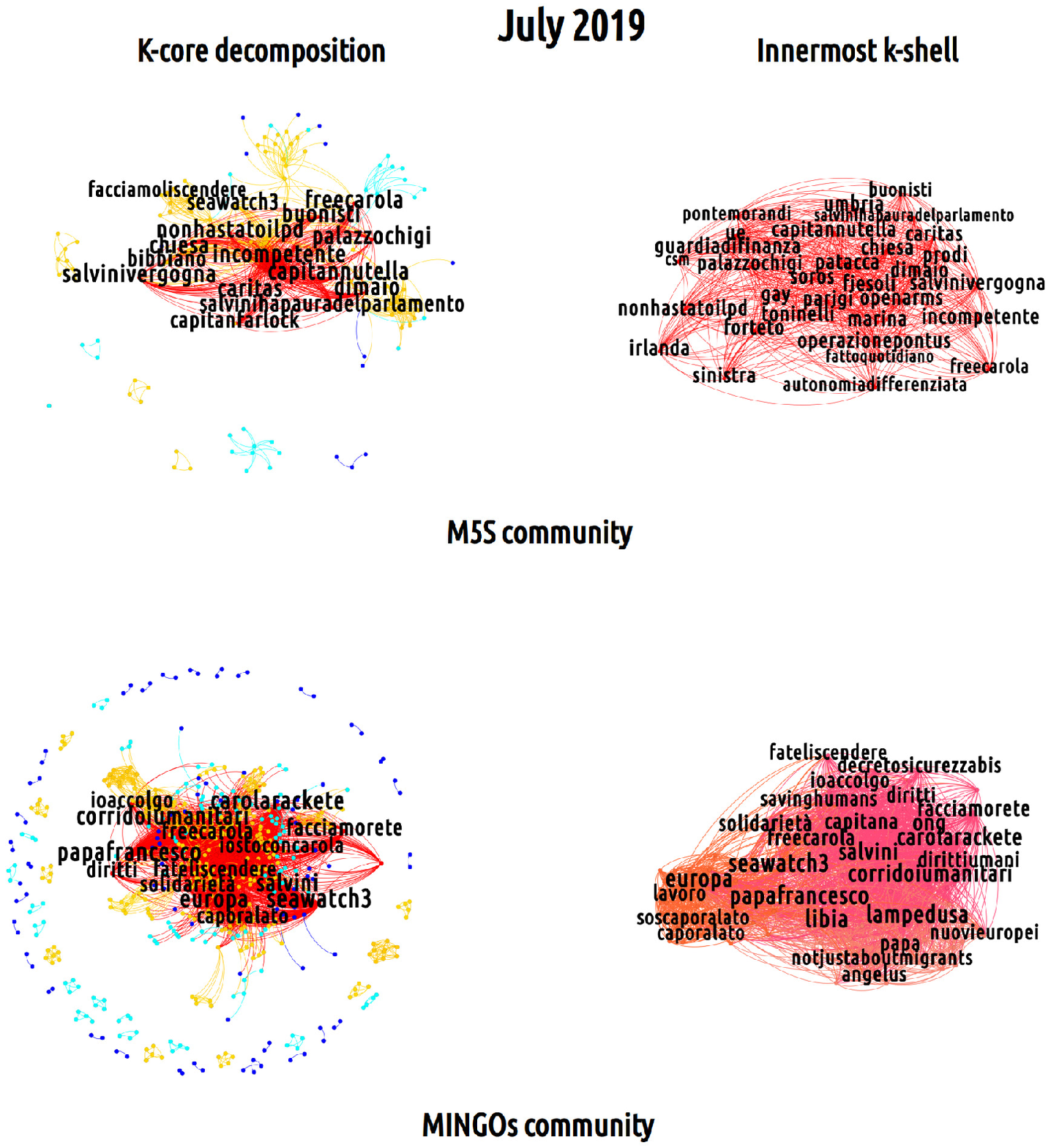}
\caption{{\bf K-core decomposition of the July 2019 semantic networks for the M5S (top left) and the MINGOs (bottom left) partisan communities.} The k-core decomposition reveals the bulk of the discussion about immigration developed by these two partisan communities: while the M5S community displays a mixed behavior towards immigration policies, as shown by the hashtags \textit{\#freecarola} and \textit{\#bibbiano}, the MINGOs innermost k-shell uncovers a strong support towards pro-migration positions, as proven by the hashtags \textit{\#ioaccolgo} (`I host'), \textit{\#iostoconcarola} (`I stand with Carola') and \textit{\#facciamorete} (`let us act as a network').}
\label{fig6}
\end{figure}

\paragraph{Semantic networks at the mesoscale: k-core decomposition.} In order to gain insight into the mesoscopic organization of semantic networks (i.e. into a less trivial dimension of networked framing), we carried out a k-core decomposition. The so-called k-core individuates a sub-graph whose nodes have a degree whose value is at least $k$. This kind of analysis partitions a network into shells as the threshold value $k$ varies. In this way, each node can be assigned a `coreness' score, depending on its level of connectedness with other vertices. Here, we have divided the distribution of k-core values into four quantiles (see \nameref{Appendix_C} for more details on this), representing the five different regions reported in Fig. \ref{fig5} and \ref{fig6} and colored from red to dark blue to indicate decreasing values of k. 

K-core decomposition has been shown to provide insightful information about the network structure in several disciplines \cite{KONG20191}. However, it does not provide any hint about the statistical significance of the recovered partition: hence, we coupled the k-core decomposition with a more traditional core-periphery decomposition; the latter one has been obtained by running the surprise minimization algorithm, a technique that has been introduced in \cite{de2019detecting}.

What emerges from the comparison between the k-core and the core-periphery decomposition is that the core overlaps to a large extent with the innermost k-shell, as the Jaccard correlation index (larger than 0.6 for all the semantic networks) confirms. Such overlap, in turn, signals that hashtags are hierarchically arranged within the Twitter discussions, i.e. that partisan communities tend to generate specific narratives that are hierarchically ordered around a finite set of thematic priorities. Analogously to what has been shown in relation to discursive dynamics developed during electoral campaigns \cite{Radicioni_2020}, also in this case discussions tend to revolve around a handful of hashtags which function as political slogans and are located in the innermost k-shell of semantic networks; on the contrary, secondary topics-, actors- or references-related hashtags are disposed in the peripheral area. A rather illustrative example is provided by the monthly-induced semantic networks of July 2019 (see Fig. \ref{fig5} and \ref{fig6}).

As shown in Fig. \ref{fig5}, the DX- and CSX-induced semantic networks display a similar topological structure: both present a tightly-connected bulk of hashtags (colored in red and orange) surrounded by a peripheral region (colored in light blue and dark blue) in which nodes are loosely inter-connected. Their red, innermost k-shell is characterized by a set of common nodes as \textit{\#carolarackete}, \textit{\#seawatch3} and \textit{\#salvini} suggesting that migration issues are specifically framed with respect to the episode of the Sea-Watch 3 and the controversial role of Matteo Salvini as the Minister of the Internal Affairs during this event. On the other hand, the analysis of the hashtags list within the innermost core (top right and top left of Fig. \ref{fig5}) provides a hint on the polarized nature of framing practices inside the two mains partisan communities: within the DX community, some of the hashtags with the greatest degree are \textit{\#portichiusi} (`close ports'), \textit{\#iostoconsalvini} (`I stand with Salvini'), \textit{\#salvininonmollare} (`Salvini don't give up'), \textit{\#arrestatecarolarackete} (`arrest Carola Rackete') and \textit{\#nonfateliscendere} (`don't let them get off'), all referring to the anti-migration slogans of the right-wing array and supporting its leader, i.e. Salvini. Conversely, the analysis of the core of the CSX-induced semantic network reveals slogans like \textit{\#portiaperti} (`open ports'), \textit{\#carolaracketelibera} (`free Carola Rackete') and \textit{\#fateliscendere} (`let them get off'), \textit{\#salvinidimettiti} (`Salvini resign'), \textit{\#ministrodellamalavita} (`ministry of the organized crime') which convey a radically opposite view, being openly against the closure of the Italian ports, sustaining clear pro-migration positions and calling for the resignation of Matteo Salvini from his position as Minister of the Internal Affairs.

For what concerns the other communities, Fig. \ref{fig6} shows the M5S- and MINGOs-induced semantic networks. While these display a hierarchical structure that is similar to that of the other two communities, their peripheral region is organized in sparser structures that are also less connected with the rest of the network, suggesting the presence of multiple framing attempts, taking place at the same time, in a rather sparse way. The red innermost k-shell of the M5S community reflects the same tension pointed out above with respect to the usage of hashtags that equally ask for the release of Carola Rackete, as \textit{\#iostoconcarola} (`I stand with Carola') and \textit{\#freecarola} but also mark a distance with the Democratic Party along the lines of hashtags endorsed also by the DX community (e.g. \textit{\#bibbiano}). Interestingly, this monthly network shows the epitomes of the rift within the governmental coalition, which passes through framing the core issue of migration in conjunction with hashtags as \textit{\#salvinivergogna} and \textit{\#salvinihapauradelparlamento} (respectively, `shame on Salvini' and `Salvini is afraid of the parliament'). Conversely, as noticed above, the Twitter discussion taking place in the MINGOs-induced semantic networks is less centered on politics. Within this community, discussions on migration and international cooperation, as shown by the hashtags \textit{\#corridoiumanitari} and \textit{\#diritti} (respectively, `humanitarian corridors' and `rights'), are more prominent. Besides, the hashtags present within the community confirm the pro-migration position endorsed by its users, as proven by the presence in the innermost k-shell of the hashtags \textit{\#ioaccolgo} (`I host'), \textit{\#iostoconcarola} (`I stand with Carola') and \textit{\#facciamorete} (`let's act as a network').

A similar configuration of framing practices that also tends to mirror fluid political alliances can be found in other monthly networks. For instance, observing the mesoscale structure of semantic networks in August 2019, it is possible to associate distinct partisanships with opposing framing practices. While there is a transversal tendency to build a connection between migration issues and the governmental crisis, partisan communities are populated by different slogans and keywords, witnessing different ways of reading this connection. On the one side, the innermost core of the CSX- and M5S-induced semantic networks reveals hashtags against Matteo Salvini’s decision to start the governmental crisis, as \textit{\#governodelfallimento}, \textit{\#legatifrega}, \textit{\#salvinitraditore} and \textit{\#salvinidimettiti} (respectively, `government of failure', `League fools you', `Salvini liar' and `Salvini traitor') thus shedding light on the progressive construction of a common ground for the future government alliance between the Democratic Party and the Five Stars Movement; on the opposite side, the DX community shows a strong endorsement for its leader and asks for new elections via hashtags as \textit{\#iostoconsalvini}, \textit{\#salvininonmollare}, \textit{\#elezionisubito} and \textit{\#vogliamovotare} (respectively, `I stand with Salvini', `Salvini don't give up', `elections now' and `we want to vote'). Interestingly, in response to these positions, the former allies from within the M5S community explicitly denounce Salvini’s (alleged) betrayal via the hashtags \textit{\#legatifrega} (`The League fools you') and \textit{\#salvinitraditore} (`Salvini traitor'). Consistently, the core of the MINGOs community does not show any specific hashtags pointing to the governmental crisis. Thus, the innermost k-shell of the semantic network continues to be populated by a set of keywords rather similar to those of the monthly semantic networks of July 2019, such as \textit{\#corridoiumanitari} (`humanitarian corridors') and \textit{\#papafrancesco} (`Pope Francis').

\section{Conclusions}

In this paper, we propose a framework to expand current analyses of online adversarial dynamics that grounds in the longitudinal exploration of the two-fold set of social and semantic relations. The application of our framework to the analysis of the Italian debate on migration issues, across the period May-November 2019, provides some interesting insights on the nature of online conflicts between political actors as well as on its multidimensional nature - a conflict to which both elites and citizens contribute by setting up partisan relationships and framing practices that evolve in a fluid fashion.

On the one hand, our results confirm those obtained in \cite{fi12100173} which, looking as we do at Twitter discussion on migration issues, find an overall disconnection between the level of engagement in online debates and the dynamics that take place on the ground. To some extent, indeed, mechanisms of online partisanship grounding online communities partly detach from those of formal political alliances: this is particularly evident in the separation between the community of the Five Stars Movement and that of the League party also during periods in which the two parties jointly shared seats within the first Conte government. Only under certain circumstances the two communities merge but, in fact, their discursive coalition is exceptional and just temporary. On the other hand, networked partisanship cannot be thought in isolation from political dynamics on the ground, as it is well demonstrated by the persistent fracture between DX and CSX communities and by the fracture within the left-wing community after the internal break of the Democratic Party.

Our results further shed light on how different social media affordances contribute to adversarial relations between online partisan communities. While these communities coalesce via retweets, they also interact, often in contentious ways, via direct mentions. Importantly, different communities leverage on technological affordances in different ways. Collective identities sustaining communities are, thus, formed in more institutional ways (mainly retweeting messages from parties and their leaders) or, more in line with a substantive criterion, re-broadcasting also messages from accounts that are more meaningfully active on migration issues. Differently, mentions are employed to construct cross-community ties that, on the one hand, soften the segregation induced by partisan endorsement while, nonetheless, often providing a means to channel antagonism. More relevantly, geometries of homophilic and cross-community ties tend to vary over time and in tight connection with relevant events on the ground - whether these are related to the issue of migration per se (as it happens in the case of contested search-and-rescue mission) or are induced by shifting political alliances (as in the case of the governmental crisis at the end of August 2019).

Closer exploration of semantic networks helped us to shed light on the more cognitive dimension of online adversarial dynamics. In the Italian case, the issue of migration seems to have provided the backbone against which multiple lines of conflict have overlapped. Along a first line, different partisan communities can be distinguished for how they approach the issue of migration: either substantively - as in the case of the MINGOs community which genuinely focuses on the complex problem of migration, often endorsing a pro-migrant point of view) - or more instrumentally, as in the case of communities shaped around political parties - which in fact leverage on migration to contrast political adversaries.

Along a second line of conflict, amongst partisan communities that discuss migration in an instrumental way, networked framing practices seem to follow more closely the fluid evolution of governmental political alliances. Regardless of the changing composition of the government, right-wing and left-wing parties remain on opposite semantic sides. This persisting arrangement mirrors at the level of networked framing practices in two ways: 1) in the case of the DX partisan community, by constructing a semantic bridge between the theme of migration and the reinforcement of internal cohesion around the figure of Matteo Salvini; 2) at the level of cross-community ties between the DX and the CSX groups, in the sustained contrast between two opposite frames on migration - one oriented towards closure (on the right), the other towards openness (on the left). What shifts (and, in fact, remains ambivalent over time) is the framing induced by the M5S community, which appears to be "semantically torn" in between its long-standing difficulty to cope with more extreme positions held by the League and its aversion for established `big parties' such as the Democratic Party. In this tension, its positioning on migration issues remains vague and, in some sense, ancillary in comparison to a much more prominent interest for discussing political dynamics.

Together, these results invite to move beyond consolidated views, often associated to the concept of political polarization, and that map down to dichotomous distinctions between positive/negative or supportive/contrary views with respect to contested issues. As such, they provide a first contribution towards a deeper, and less obvious, understanding of how oppositional and partisan conflict occurs in ``non perfectly polarized contexts'' where multiple ideological poles are present and often stand in complex inter-relations \cite{fi12100173, doi:10.1177/0163443719876541}.

Above and beyond specifying the social and semantic peculiarities of the considered adversarial debate, our approach innovates the study of online political discussions in two main ways. On the one hand, it grounds semantic analysis within users' behaviors by implementing a method, rooted in statistical theory, that guarantees that our inference of socio-semantic structures is not biased by any unsupported assumption about missing information. On the other hand, our operational approach represents an unsupervised algorithm for detecting partisan communities and semantic networks. In fact, the network approach proposed in this paper provides a method for extracting relevant political information from a Twitter discussion without relying on any pre-existent information on users or media contents. As a consequence, our method is suitable for application to any Twitter discussion.

\section{Acknowledgements}
F.S. and T.S. acknowledge support from the European Project SoBigData++ (GA. 871042). F.S. also acknowledges support from the Italian `Programma di Attività Integrata' (PAI) project `TOols for Fighting FakEs' (TOFFE), funded by IMT School for Advanced Studies Lucca. E.P. acknowledges support from the project `I-Polhys - Investigating Polarization in Hybrid Media Systems' funded by the Italian Ministry of University and Research within the PRIN 2017 framework (Research Projects of Relevant National Interest for the year 2017; project code: 20175HFEB3). 

\bibliographystyle{unsrt}
\bibliography{MigrantsPaper}

\newpage

\section*{Appendix}





\paragraph*{S1 Appendix.}\label{app:events} {\bf Overview of events related to the Italian debate on migration.} Since the beginning of 2014, when the numbers of migrants attempting to enter the European Union massively increased, the refugees crisis is a hot-topic widely discussed on the European mainstream media \cite{berry2016press}. The prominence of this topic has been increased also due to the rise of European political parties fuelling xenophobic or anti-migration policies or narratives that have built a rhetorical construction of immigrants and refugees as a dangerous threat for national security.

In June 2018, the two (initially competing) parties emerged as winners after the Election day, i.e. the Five Stars Movement and the League, formed a new government. In the self-proclaimed `government of change', the League party, albeit the minority partner in the coalition, successfully succeeded in influencing the political agenda of the government on several topics at the center of its political project, e.g. the migration policies. One of most debated decisions taken by Matteo Salvini was the closure of the Italian ports to the NGOs boats rescuing migrants in the Mediterranean Sea. This political initiative not only took the tangible forms of entry bans into Italian ports or seizing of several rescue vessels but also in the approval of two decrees on security - the second one called Second Security Act (`Decreto Sicurezza Bis') was approved in June 2019 - the increased the difficulties, for asylum seekers, to request the residence permits issued for humanitarian reasons and introduced fines for NGOs active in migrants rescuing activities. On 19 August, approximately 14 months after its formation, the governement fell upon the initiative of Matteo Salvini that submitted a no-confidence motion against the Prime Minister Giuseppe Conte. Few weeks later, the M5S formed a new government with the Democratic Party (and other minor forces of the Italian left-wing parties) while the League went back to the opposition. The new government immediately presented a different attitude towards the NGOs involved in search-and-rescue operations: on 26 October, the new Minister of the Internal Affairs, Luciana Lamorgese, meet the organizations to discuss their activities.

The other major political events of interest for our analysis are 1) the European election on 26 May 2019 and 2) the entering of the rescue vessel Sea-Watch 3 into the Italian territorial waters, without permission, at end of June 2019. The former one is a significant electoral event since the League became the most voted Italian party, with 34.1\% of the vote share (that reversed the political result of the 2018 general elections, after which the Five Stars Movement emerged with a similar percentage of preferences); the latter one is one of the most debated events about migration before the fall of the government: after two weeks sailing, on 29 June 2019, the rescue vessel Sea-Watch 3 entered the Italian territorial waters and its captain Carola Rackete was arrested. Matteo Salvini, the back-then Minister of the Interior, accused Rackete of hitting an Italian patrol boat which tried to intercept the Sea-Watch 3 before docking. At the same time, Rackete was under investigation by Italian authorities for alleged criminal activities concerning undocumented activities in search-and-rescue operations. After a brief detention under house arrest, she was released by an Italian court ruling that she acted to protect the safety of the passengers.

Table \ref{tab:events} provides a brief overview of the most relevant political and mediatic events concerning the Italian Twittersphere discussion about migration. (see also \cite{fi12100173} for an overview of the Italian discussion on migration issues across the period August 2018-July 2019).

\paragraph*{S1 Table.} {\bf Overview of the most relevant political and mediatic events for what concerns the Italian Twittersphere discussion about migration.}

\begin{table}[!ht]
\centering
\caption{{\bf Overview of the most relevant political and mediatic events concerning the Italian Twittersphere discussion about migration.}}
\begin{tabular}{l|p{14cm}}
\hline
\textbf{Date} & \textbf{Event} \\
\hline
\hline
May 18$^\text{th}$ & Italian authorities seize the rescue vessel Sea-Watch 3 after 47 migrants have been rescued. \\
\hline
May 26$^\text{th}$ & 2019 European elections. \\
\hline
\hline
Jun 1$^\text{th}$ & Sea-Watch 3 ends its seizure period. \\
\hline
Jun 7$^\text{th}$ & The European Council adopts a partial negotiation position about migrants repatriation and new guidelines for security.\\
\hline
Jun 11$^\text{th}$ & The Italian Council of Ministers approves the `Decreto Sicurezza Bis' (`Second Security Act') against illegal immigration and NGOs.\\
\hline
Jun 29$^\text{th}$ & After two weeks sailing, Sea-Watch 3 enters the Italian territorial waters without any formal permission; its captain Carola Rackete is arrested.\\
\hline
\hline
Jul $19^\text{th}$ & After her freeing, Carola Rackete goes back to Germany.\\
\hline
Jul $24^\text{th}$ & Italian Parliament approves the `Decreto Sicurezza Bis' (`Second Security Act').\\
\hline
\hline
Aug 1$^\text{st}$ & The rescue vessel Open Arms rescues 52 migrants.\\
\hline
Aug 9$^\text{th}$ & The League party submits a no-confidence motion against the Prime Minister Giuseppe Conte. The Italian government crisis starts.\\
\hline
Aug 19$^\text{th}$ & After the refusal of the Italian government, the Spanish Prime Minister Pedro Sanchez allows Open Arms to disembark migrants in Mallorca.\\
\hline
\hline
Sep 2$^\text{nd}$ & Italian authorities seize the rescue vessel Eleonore.\\
\hline
Sep 5$^\text{th}$ & The Italian government crisis ends with the Conte-bis cabinet swearing in.\\
\hline
Sep 14$^\text{th}$ & The rescue vessel Ocean Viking is allowed to disembark 82 rescued migrants in Lampedusa.\\
\hline
\hline
Oct 26$^\text{th}$ & The NGOs involved in search-and-rescue operations at sea meet the new Minister of the Internal Affairs Luciana Lamorgese.\\
\hline
\hline
Nov 1$^\text{st}$ & After seven days sailing, the rescue vessel Alan Kurdi is allowed to dock in Taranto.\\
\hline
\end{tabular}
\label{tab:events}
\end{table}
  
\paragraph*{S2 Appendix.}\label{app:hashtag_descriptions} {\bf Description of the main hashtags.} In our analysis, hashtags might be obscure to a reader who is not familiar with the Italian language and with societal and political references related with them. For this reason, we provide a translation of the main Italian hashtags along with a brief description of the Italian political and societal context.

\paragraph*{S2 Table.} \textbf{Brief description of the main hashtags present in our analysis.}

\begin{table}[!ht]
\centering
\caption{{\bf Brief description of the main hashtags present in our analysis.}}
\begin{tabular}{p{4.5cm}|p{13cm}}
\hline
\textbf{Hashtag} & \textbf{Translation and context} \\
\hline
\hline
\textit{\#salvini} & Matteo Salvini is the leader of the right-wing party called the League. He was Minister of the Interior until August 2019 when the League submitted a no-confidence motion against the Prime Minister Giuseppe Conte.\\
\hline
\textit{\#oceanviking}, \textit{\#seawatch3}, \textit{\#carolarackete}, \textit{\#lampedusa}, \textit{\#marejonio} & Ocean Viking, Sea Watch 3 and Mare Jonio are names of NGOs (called \textit{ONG} in Italian) rescue boats whose activity is that of providing the first aid to migrants at sea. Lampedusa is an Italian island, approximately at the same distance from Malta, Sicily and the coasts of Tunisia, where rescued migrants have been often disembarked. Carola Rackete is the captain of Sea-Watch 3 who forced the docking ban entering the Italian territorial waters without any formal permission.\\
\hline
\textit{\#blocconavale} & Literally, `ship block'. Ship blocks represented the strategy used by the Minister of the Internal Affairs Matteo Salvini to forbid NGOs to disembark migrants in Italy.\\
\hline
\textit{\#portichiusi}, \textit{\#portiaperti} & Respectively, `closed ports' and `open ports'. The first hashtag was used as a support of the Matteo Salvini's political slogan which summarizes his political view about migration policies. The second one has been invented by political adversaries as a reversed version of the initial slogan.\\
\hline
\textit{\#bibbiano} & Bibbiano is a small municipality in Emilia Romagna, a northern Italian region, where a judicial case involved several center-left political actors, as the major of the city, as part of a criminal business for allegedly brainwashing vulnerable children. Although this event is not directly connected with immigration issues, the demand for justice for case victims has been one of the topics of the League and Brothers of Italy against the center-left coalition.\\
\hline
\textit{\#fateliscendere}, \textit{\#iostoconcarola}, \textit{\#salvinivergogna}, \textit{\#salvinihapauradelparlamento} \textit{\#salvinidimettiti}, \textit{\#ministrodellamalavita} & Respectively, `let them get off', `I stand with Carola', `shame on Salvini', `Salvini is afraid of the parliament', `Salvini resign' and `ministry of the organized crime'. These hashtags were used during the Sea-Watch 3 episode by the CSX, M5S and MINGOs communities to either support Matteo Salvini or Carola Rackete (depending on the position about migration).\\
\hline
\textit{\#nonfateliscendere}, \textit{\#iostoconsalvini}, \textit{\#salvininonmollare}, \textit{\#arrestatecarolarackete} & Respectively, `don't let them get off', `I stand with Salvini', `Salvini don't give up' and `arrest Carola Rackete'. These hashtags were used by the DX community to show support towards the governmental policies on migration promoted by the Minister of the Internal Affairs at the time, i.e. Matteo Salvini, during the Sea-Watch 3 case.\\
\hline
\textit{\#clandestini} & Literally, `illegal immigrants'. A term mostly used by Italian right- and far-wing parties to call migrants who land in the Italian coasts.\\
\hline
\textit{\#facciamorete} & Literally, `let's act as a network'. A political slogan used to call for an antifascist grassroots movement dealing with various civic issues and born to counteract the migration policies of Matteo Salvini.\\
\hline
\textit{\#governodelfallimento}, \textit{\#legatifrega}, \textit{\#salvinibugiardo}, \textit{\#salvinitraditore} & Respectively, `government of failure', `League fools you', `Salvini liar' and `Salvini traitor'. These hashtags were used by CSX and M5S community to show dissatisfaction towards the behavior of the League and its political leader, leading to the government crisis. The slogan `government of failure' is a pun linked to the self-proclaimed `government of change' of the Five Stars Movement and the League.\\
\hline
\textit{\#crisidigoverno}, \textit{governoconte2}  & Respectively, `government crisis' and `government Conte 2'. These two hashtags were used by M5S community to refer to the political crisis and the newly formed government of the Prime Minister Giuseppe Conte with the support of the Five Stars Movement and the Democratic Party.\\
\hline
\textit{\#elezionisubito}, \textit{\#vogliamovotare} & Respectively, `elections now' and `we want to vote'. These hashtags were used by the DX community to show their dissatisfaction towards the agreement between the Five Stars Movement and the Democratic Party to form a new government after the crisis.\\
\hline
\textit{\#dirittiumani}, \textit{\#inclusione}, \textit{\#corridoiumanitari}, \textit{\#ioaccolgo}, \textit{\#giornatamondialedelrifugiato} & Respectively, `human rights', `inclusion', `humanitarian corridors', `I host' and `world refugee day'. These hashtags were used in the CSX and MINGOs communities to discuss about migration issues as search-and-rescue activities, inclusion and human rights at a national and European level.\\
\hline
\textit{\#decretosicurezzabis} & Literally, `second security act'. This act represents a revision, proposed by the League party, of Italian migration policies introducing fines for NGOs search-and-rescue activities and stricter rules for requesting the residence permits issued for humanitarian reasons.\\
\hline
\end{tabular}
\label{tab:hashtag_descriptions}
\end{table}

\paragraph*{S3 Appendix.}\label{app:inferring} {\bf Bipartite networks projection and validation.} This section provides a brief overview of the algorithm we have implemented to project our bipartite networks on a single layer (be it the one of verified users or the one of hashtags). Generally speaking, this procedure outputs a monopartite projection by linking any two nodes, belonging to the same layer, if the number of their common neighbors is statistically significant; it can be summarized into three steps.

First, a measure quantifying the degree of similarity between two nodes is needed. Given any two nodes $\alpha$ and $\beta$ of the same layer $\bot$, their similarity is provided by the total number of co-occurrences, i.e. the number of common neighbors $V_{\alpha\beta}^*$, computable as $V_{\alpha\beta}^*=\sum_{j=1}^{N_\top}V_{\alpha\beta}^j=\sum_{j=1}^{N_\top}m_{\alpha j}m_{\beta j}$. The term $V_{\alpha\beta}^j$=$m_{\alpha j}m_{\beta j}$ denotes the `single' common neighbor, defined by the nodes $\alpha$ and $\beta$ with $j$ belonging to the opposite layer; its value is 1 if nodes $\alpha$ and $\beta$ share node $j$ as a common neighbor and 0 otherwise.

Second, the statistical significance of any two nodes similarity needs to be quantified. To this aim, observations have to be compared against a proper null model that can be defined within the mathematical framework of the so-called Exponential Random Graphs. This framework is based on a very general principle rooted in statistical physics \cite{SquartiniCimini2019}, prescribing to employ the conservative benchmarks that are induced by the maximization of Shannon entropy. In mathematical notation, given the ensemble $\mathcal{M}$ of networks and the probability $P(\mathbf{M})$ of occurrence of a network $\mathbf{M}\in\mathcal{M}$, the Shannon entropy is

\begin{eqnarray}
\label{eq:entropy}
S=-\sum_{\mathbf{M}\in\mathcal{M}} P(\mathbf{M})\ln P(\mathbf{M})
\end{eqnarray}
where the sum runs over the set of all possible bipartite graphs with, respectively, $N_\top$ nodes on the $\top$ layer and $N_\bot$ nodes on the $\bot$ layer - as the real network system $\mathbf{M^{*}}$. As the entropy-maximization procedure is carried out in a constrained framework, let us define the constraints of the Bipartite Configuration Model (BiCM) \cite{Saracco_2015}, i.e. the null model adopted in the present paper. In this specific model, the ensemble average of the degrees of users and hashtags (i.e. $k_i^*=\sum_\alpha m_{i\alpha},\:\forall\:i$ and $h_\alpha^*=\sum_i m_{i\alpha},\:\forall\:\alpha$, respectively) are considered as fixed. Upon introducing the Lagrange multipliers $\boldsymbol{\theta}$ and $\boldsymbol{\eta}$ to enforce the proper constraints and $\psi$ to ensure the normalization of the probability, the recipe prescribes to maximize the Lagrangian function

\begin{eqnarray}
\mathcal{L}=S-\psi\left[1-\sum_{\mathbf{M}\in\mathcal{M}}P(\mathbf{M})\right]-\sum_{i=1}^{N_\top}\theta_i[k_i^*-\langle k_i\rangle]-\sum_{\alpha=1}^{N_\bot}\eta_\alpha[h_\alpha^*-\langle h_\alpha\rangle]
\end{eqnarray}

with respect to $P(\mathbf{M})$. This leads to:

\begin{eqnarray}
P(\mathbf{M}|\boldsymbol{\theta},\boldsymbol{\eta})=\prod_{i=1}^{N_\top}\prod_{\alpha=1}^{N_\bot}p_{i\alpha}^{m_{i\alpha}}(1-p_{i\alpha})^{1-m_{i\alpha}}
\end{eqnarray}

where $x_i\equiv e^{-\theta_i}$, $y_\alpha\equiv e^{-\eta_\alpha}$ and the quantity $p_{i\alpha}\equiv\frac{x_iy_\alpha}{1+x_iy_\alpha}$ is the probability that user $i$ and hashtag $\alpha$ are connected (i.e. that $m_{i\alpha}=1$). 

Links independence under the BiCM implies that 1) the presence of a co-occurrence (i.e. $m_{\alpha j}m_{\beta i}=1$) can be described as the outcome of a Bernoulli trial whose probability reads $f_\text{Ber}(m_{\alpha j}m_{\beta i}=1)=p_{\alpha j}p_{\beta j}$ and that 2) the term $V_{\alpha\beta}$ is the sum of independent Bernoulli trials, each one characterized by a different probability. The behavior of such a random variable is described by the Poisson-Binomial (PB) distribution. Thus, quantifying the statistical significance of the similarity of nodes $\alpha$ and $\beta$ amounts at computing

\begin{equation}
\text{p-value}(V_{\alpha\beta}^{*})=\sum_{V_{\alpha\beta}\geq V_{\alpha\beta}^{*}} f_\text{PB}(V_{\alpha\beta});
\end{equation}
this procedure is repeated for each pair of nodes, hence obtaining $\binom{N_\bot}{2}$ p-values. 

Third, in order to understand which p-values are significant, a validation procedure for testing simultaneously multiple hypotheses is needed. The choice of the present paper has been directed towards the False Discovery Rate (FDR) \cite{benjamini1995controlling} which prescribes to, first of all, sort the $\binom{N_\bot}{2}$ p-values in increasing order, i.e. p-value$_{1}\leq\ldots\leq$p-value$_{n}$, and, then, identify the largest integer $\hat{i}$ satisfying the condition $\mbox{p-value}_{\hat{i}}\leq\frac{\hat{i}t}{n}$, where $t$ represents the single-test significance level (in our case, it is set to 0.01). All p-values that are less than, or equal to, $\mbox{p-value}_{\hat{i}}$ are, thus, kept, meaning that all the corresponding node pairs are linked in the resulting monopartite projection.

\paragraph*{S4 Appendix.}\label{Appendix_C} {\bf Analysis of mesoscale network structures.} The Louvain algorithm \cite{Blondel_2008} has been run on the monopartite projections of the user networks to detect the presence of communities. This algorithm works by searching for the partition attaining the maximum value of the modularity function $Q$, i.e.

\begin{eqnarray}
Q=\frac{1}{2L}\sum_{i,j}\left[a_{ij}-\frac{k_ik_j}{2L}\right]\delta_{c_i,c_j}
\end{eqnarray}

a score function measuring the optimality of a given partition, by comparing the empirical pattern of interconnections with the one predicted by a properly-defined benchmark model. In the formula above, $a_{ij}$ is the generic entry of the network adjacency matrix $\mathbf{A}$, the factor $\frac{k_ik_j}{2L}$ is the probability that nodes $i$ and $j$ establish a connection according to the Chung-Lu model, $c$ is the $N$-dimensional vector encoding the information carried by a given partition (the $i$-th component, $c_i$, denotes the module to which node $i$ is assigned) and the Kronecker delta $\delta_{c_i,c_j}$ ensures that only the nodes within the same modules provide a positive contribution to the sum. For the present analysis, as already pointed out in \cite{Saracco_2019}, a reshuffling procedure has been applied to overcome the dependence of the original algorithm on the order of the nodes taken as input. In order to prevent the modularity function from being stuck in a local maximum, the Louvain algorithm is repeated 1.000 times which is a number of runs ensuring the algorithm stability.

Regarding the k-core and the core-periphery decompositions, we briefly describe our approach below. While a proper core-periphery detection can be carried out by adopting the method proposed in \cite{de2019detecting} and prescribing to search for the network partition minimizing the surprise score function, the presence of a densely connected subgraph can be inspected in a simpler way, i.e. via the k-core decomposition. A k-core is a subgraph whose nodes have a degree whose value is at least $k$ and can be revealed in an iterative fashion (i.e. deleting nodes with degree less than $k$). A node is said to have `coreness' $c$ if it belongs to a $c$-core but not to any $(c+1)$-core.

An interesting phenomenon that is observed when analysing monthly semantic networks is the co-existence of different types of mesoscopic structures. As pointed out in other works \cite{malvestio2020interplay}, communities are found to co-exist with k-shells in several systems: in particular, the innermost k-shells is frequently sub-divided into (a small number) of communities further partitioning it. These sub-communities play an important role: within our partisan communities-induced semantic networks, this role has been interpreted as a natural division of the most debated topics animating the same `global' discussion. Running the Louvain community detection algorithm on the innermost shell (reported on the right in Fig. \ref{fig5} and Fig. \ref{fig6}) indeed reveals the existence of these sub-communities. For instance, the core of the DX community displays two distinct sub-communities: one of the clusters concerns a general discussion on the regulation of the migration flows and the behavior of Italian institutions in handling
migrants - e.g. hashtags like \textit{\#braccianti} (`day labourer'), \textit{\#permessi} (`permits') and \textit{\#confisca} (`confiscation' - appear); the second one concerns the Sea-Watch 3 crisis and the opposition to its entrance into the Italian ports (e.g. hashtags like \textit{\#blocconavale} (`ship block'), \textit{\#iostoconsalvini} (`I stand with Salvini') and \textit{\#salvininonmollare} (`Salvini don't give up') appear).

Remarkably, the nodes connecting the thematic clusters act as `bridges' and represents more general hashtags about migration policies (e.g. \textit{\#salvini} and \textit{\#libia}). In this way, the picture of the list of hashtags with largest betweenness centrality can be further refined upon looking at the innermost k-shells. As an example, let us consider the DX-induced semantic network for the month of July 2019: the keywords playing the role of bridges read \textit{\#salvini}, \textit{\#capitana} (an untranslatable hashtags referring to Carola Rackete as the captain of the rescue vessel Sea Watch 3) and \textit{\#libia}, a signal that these hashtags are included into several discussions within this partisan community.

\paragraph*{S5 Appendix.}\label{Appendix_D} {\bf Computing the polarization of non-verified users.} Once the communities of verified users are detected, the procedure to build partisan communities ends with the `inclusion', within these communities, of non-verified users. In order to do that, the full retweeting network (with all the Twitter users involved in the discussion throughout the entire observation period) is, now, considered. In this network, the information about the community affiliation of the verified users is preserved via a \emph{community label}. A measure of the distribution of the community label of the neighbors is, then, assigned to each non-verified user. This measure is called \emph{polarization index} and determines the balance of the interactions of each non-verified user, in the retweeting network - namely how the portion of interactions of the non-verified users is distributed `towards' each community of verified users.

Let define $C_{c}$ (where $c$ represents the community label) as the set of users belonging to community $c$ with whom the non-verified user $\alpha$ has interacted and $N_{\alpha}$ as the set of neighbors of $\alpha$ with whom he interacted with. The \emph{polarization index} for each non-verified user $\alpha$ is defined as:

\begin{equation}
\rho_{\alpha}=\max_c\{I_{\alpha c}\}
\label{eq:polarization}
\end{equation}

where:

\begin{equation}
I_{\alpha c}=\frac{|C_c\cap N_\alpha|}{|N_\alpha|}.
\end{equation}

As shown in \cite{Saracco_2019}, the polarization index reveals how unbalanced the distribution of retweets, thus providing a clear indication of the target community of each non-verified user. To be sure we include only non-verified users with large values of polarization index, we, first, define a \emph{polarized user} as a non-verified user with a value of the polarization index greater than, or equal to, 0.9. All polarized users are, thus, included within the corresponding community of verified users.

After this first step, we have inferred the orientation of the remaining, non-verified users by propagating the tags of each partisan community obtained for the verified users. A label propagation algorithm, as the one proposed in \cite{raghavan2007near}, has been run on the retweeting network: this algorithm implements the idea that each node in the retweeting network joins the same community the majority of its neighbors belongs to. Each verified user is initialized with a unique community label while all the other users (including the polarized ones) have no label; then, these labels propagate through the network until densely connected groups of nodes reach a consensus. In case no majority is found, the algorithm randomly removes a link and re-evaluates the labels \cite{raghavan2007near}. Due to the stochasticity introduced by this latter step, the label propagation algorithm is repeated 1.000 times. In fact, although this algorithm can modify the label assigned to a non-verified user by the polarization index, labels are still stable after a large number of runs.  At the end of such a process, each non-verified user is assigned the community label met with higher frequency across all iterations. After this last step, more than $90\%$ of the Twitter users are assigned to one of the main partisan communities.

The stability of the results of such an algorithm has been already proven in \cite{Caldarelli2020c} whose authors compare several approaches in order to check the presence of dependencies on the different sets of assumptions. In particular, authors of \cite{Caldarelli2020c} consider two algorithms, i.e. the Louvain community detection  algorithm and the label propagation algorithm proposed in \cite{raghavan2007near}, and apply them to three different representations of the retweeting network: the original weighted network of retweets, its unweighted version and the validated projection obtained via the procedure proposed in \cite{Becatti2019}. The Variation of Information (VI) (i.e., an information theoretic criterion for comparing any two partitions, defined in \cite{Meila2003}) is then used to capture the distance among the different partitions. The greatest discrepancies are found between the results of the Louvain algorithm and those of the label propagation algorithm, regardless of the network. In fact, while relatively large VI values are found across the partitions detected by Louvain on the different network representations, the label propagation induces lower VI values across the same partitions.

The aforementioned stability is probably due to several reasons. First, the seeds of the label propagation (i.e., the labels which have fixed values) are themselves extremely stable, as a result of the strict validation procedure. In this sense, the rationale upon which the projection algorithm is based seems to be particularly sound: any two verified users interacting with the same non-verified users are indeed perceived as similar. This interpretation is based on two essential features of the system, i.e. the strong modular structure of the network and the peculiar activity of verified users, much more focused on the production of original messages than on retweeting. Second, due to the adversarial dynamics observed in the present discussion, the label propagation procedure seems not to be sensitive to the topological variation induced by different representations of the retweeting network: indeed, the neighbours of a node are more likely to belong to its community irrespectively of the relative importance of the information carried by the weight.

\paragraph*{S6 Appendix.}\label{Appendix_E} {\bf Computing the h-index of Twitter users.}A useful tool to measure the influence of a given Twitter user, within a discussion, is the well-known Hirsch index (or h-index), widely used to measure the impact of research activity. This index has been also employed for quantifying the ability of Twitter users to produce influential contents via the number of retweets of each of their messages. In \cite{Brexit2017}, h-inde has been used to detect the most influential users during the Brexit debate on Twitter: similarly, we employed it to derive a ranking list of the verified users in the Italian discussion on migration issues. As Tab. \ref{tab:h_index} shows, our results further confirm the overall strong personalization of the debate on migration: in fact, Matteo Salvini's Twitter account (\textit{@matteosalvinimi}) has the highest h-index, amounting at more than twice the second ranked value. The presence of several verified accounts belonging to media organizations, such as \textit{@repubblica} and \textit{@Avvenire\_Nei}, is another indicator of how the construction of partisan collective identities relates to the activity of these accounts, e.g. in the CSX and the MINGOs communities.

\paragraph*{S6 Table.} {\bf List of the first ten verified users with the highest values of h-index (computed by considering the activity throughout the entire observation period).}

\begin{table}[ht!]
\centering
\caption{{\bf List of the first ten verified users with the highest values of h-index (computed by considering the activity throughout the entire observation period).}}
\begin{tabular}{l|c|c} 
\hline
Twitter user (screen name) & h-index & Discursive community \\ [0.5ex] 
\hline
matteosalvinimi & 231 & DX  \\ [0.5ex]
GiorgiaMeloni & 95 & DX  \\ [0.5ex]
LegaSalvini & 52 & DX \\ [0.5ex]
Linkiesta & 50 & CSX \\ [0.5ex]
Capezzone & 50 & DX \\ [0.5ex]
fattoquotidiano & 48 & M5S \\ [0.5ex]
Agenzia\_Ansa & 43 & CSX \\ [0.5ex]
Avvenire\_Nei & 43 & MINGOs \\ [0.5ex]
repubblica & 43 & CSX \\ [0.5ex]
CarloCalenda & 40 & CSX \\ [0.5ex]
\hline
\end{tabular}
\label{tab:h_index}
\end{table}

\end{document}